\documentstyle[twoside,epsfig,cite]{hsproc}
\def\runauthor{Paul Newman}
\def\shorttitle{Diffractive Phenomena at HERA}
\def\be{\begin{equation}}
\def\ee{\end{equation}}
\def\bea{\begin{eqnarray}}
\def\eea{\end{eqnarray}}

\newcommand{\av}[1]{\mbox{$ \langle #1 \rangle $}}

\newcommand{\gapprox}{\stackrel{>}{_{\sim}}}
\newcommand{\lapprox}{\stackrel{<}{_{\sim}}}
\newcommand{\pom}{\rm I\!P}
\newcommand{\reg}{\rm I\!R}
\newcommand{\pt}{p_{_T}}
\newcommand{\mx}{M_{_X}}
\newcommand{\my}{M_{_Y}}
\newcommand{\alphapom}{\alpha_{_{\rm I\!P}}}
\newcommand{\alphareg}{\alpha_{_{\rm I\!R}}}

\newcommand{\xpom}{x_{_{\pom}}}
\newcommand{\xl}{z}
\newcommand{\ptj}{p_{_T}^{\rm jet}}
\newcommand{\zpom}{z_{_{\rm \pom}}}

\newcommand{\zpomj}{z_{_{\rm \pom}}^{\rm jets}}
\newcommand{\xgamj}{x_{\gamma}^{\rm jets}}
\def\etalk{{ et al., }}

\begin{document}
\title{Diffractive Phenomena at HERA \footnote{Based on an invited
talk at the `Hadron Structure 98' Conference, Stara Lesna, Slovakia.}}
\author{Paul Newman\footnote{Supported by the 
UK Particle Physics and Astronomy Research Council 
(PPARC).}\hspace{0.1cm} (\email{prn@hep.ph.bham.ac.uk}) \\
for the H1 and ZEUS Collaborations.}
{School of Physics, University of Birmingham, B15 2TT, UK}
%
\abstract{Highlights selected from 
recent measurements of colour-singlet exchange processes at 
HERA are summarised. Particular emphasis is placed on energy
and momentum transfer dependences and the decomposition of the data 
into an expansion in Regge trajectories. The latest results in vector meson 
helicity analysis, partonic descriptions of hard diffraction and 
diffractive dijet production are also covered.}
%


\section{Introduction}

When discussing the success of the standard model, it is often overlooked
that the bulk of hadronic cross sections remain rather poorly understood
within the gauge theory of the strong interaction, quantum chromodynamics
(QCD). Before the advent of QCD, many aspects of hadronic interactions
were understood in the framework of Regge phenomenology, which is based on
the most general properties of the scattering matrix. In Regge models, 
diffractive (elastic, dissociative and, via the optical theorem, total) 
cross sections are
well described \cite{diff:review} 
at high energy by the exchange of the leading vacuum
singularity or {\em pomeron}. 
The relationship between Regge
asymptotics and QCD is far from clear. 

The lepton beam at HERA 
can be considered as a prolific source of
high energy real and virtual photons, such that diffractive scattering in 
the $\gamma^{(*)} p$ system can be studied. 
A variety of hard scales can be introduced to the problem, such that
HERA provides an excellent opportunity to study diffraction in regions in which
perturbative QCD may be applicable. 
This leads to the exciting possibility of
gaining an understanding of diffraction and the pomeron at the level of 
parton dynamics and of studying the transition between perturbatively 
calculable and perturbatively incalculable strong interactions.
The H1 and ZEUS experiments have produced many new and innovative measurements
that have already led to improvements in the QCD description of
diffraction. This contribution summarises the most recent developments in
colour-singlet exchange physics at HERA. A summary of older results can be
found, for example, in \cite{chicago}.

\begin{figure}[htb]
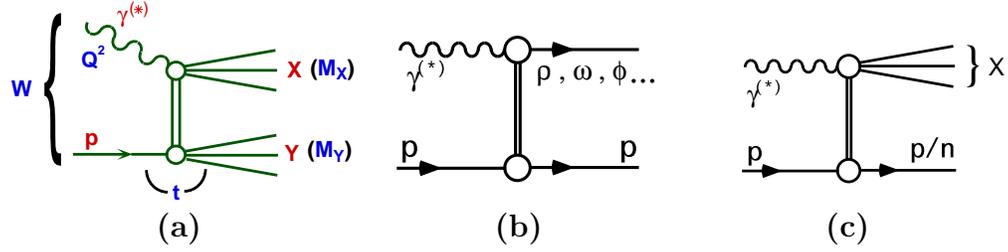
 \unitlength 1mm
 \begin{center}
   \begin{picture}(150,22)
     \put(0,0){\epsfig{file=generalhera.epsf,width=0.35\textwidth}}
     \put(51,3){\epsfig{file=EL.epsf,width=0.27\textwidth}}
     \put(97,3){\epsfig{file=GD.epsf,width=0.27\textwidth}}
     \put(20,-4){\large{\bf{(a)}}}
     \put(65,-4){\large{\bf{(b)}}}
     \put(109,-4){\large{\bf{(c)}}}
   \end{picture}
 \end{center}
  \caption  {(a) Illustration of the generic process
$\gamma^{(*)} p \rightarrow XY$. (b) Exclusive vector meson production
$\gamma^{(*)} p \rightarrow Vp$. (c) Semi-inclusive colour-singlet exchange
$\gamma^{(*)} p \rightarrow Xp$ and $\gamma^{(*)} p \rightarrow Xn$.}
  \label{generalhera}
\end{figure}

The generic colour-singlet exchange 
process $\gamma^{(*)} p \rightarrow XY$ at HERA is
illustrated in figure~\ref{generalhera}a. The 
hadronic final state consists of two distinct systems $X$ and $Y$, with $Y$ 
being the closer to the outgoing proton direction. 
The processes principally considered here are exclusive vector meson production
$\gamma^{(*)} p \rightarrow V p$ where $V = \rho$, $\omega$, $\phi$ \ldots 
(fig~\ref{generalhera}b) and 
semi-inclusive
virtual photon dissociation $\gamma^{(*)} p \rightarrow X p$, for all
systems $X$ (fig~\ref{generalhera}c). 
The semi-inclusive charge exchange reaction 
$\gamma^* p \rightarrow X n$ is also discussed.
Measurements of processes in which the proton dissociates to higher mass
systems $Y$ are described
elsewhere \cite{alexei,zeus:hight,zeus:gprho,mx:gammap}.

\section{Kinematic Variables}
\label{kine}

With $q$ and $P$ denoting the 4-vectors of the incoming photon and proton
respectively, the standard
kinematic variables 
\begin{eqnarray}
\label{eq1}
Q^2 \equiv -q^2 \hspace{3.0cm} x \equiv \frac{Q^2}{2 q.p} \hspace{3.0cm} 
W^2 \equiv (q + P)^2 \ ,
\end{eqnarray}
are defined. In the context of figure~\ref{generalhera}a,
with $p_{_X}$ and $p_{_Y}$ 
representing the 4-vectors of the final state systems $X$ and $Y$
respectively, three further variables are introduced;
\begin{eqnarray}
\label{eq2}
\mx^2 \equiv p_{_X}^2 \hspace{3.0cm} \my^2 \equiv p_{_Y}^2 
\hspace{3.0cm} 
t \equiv (P - p_{_Y})^2 \ ,
\end{eqnarray}
where $t$ is the squared four-momentum transferred between the photon and the 
proton.
The hermetic nature of the H1 and ZEUS detectors makes it possible to measure
all kinematic variables defined in equations~\ref{eq1} and~\ref{eq2} in
many circumstances. Where both $Q^2$ and $|t|$ are small, colour singlet
exchange at HERA follows a similar pattern to that
in soft hadron-hadron 
physics \cite{rho:gammap,zeus:gprho,mx:gammap,zeus:mx}. 
The region of large $|t|$ is believed to be a particularly good filter for
hard diffractive processes in which the pomeron itself may be perturbatively
calculable in terms of the exchange of a pair of
gluons from the proton in a net colour-singlet 
configuration \cite{low:nussinov}.
First measurements in
this region are starting to appear \cite{zeus:hight,hight}.
This document is principally concerned with
the kinematic region in which $|t|$ is small and a large value of $Q^2$
sets a hard scale. 

For virtual photon dissociation $\gamma^* p \rightarrow X p$, 
the kinematics are
usually expressed in terms of the variables
\begin{eqnarray}
\label{eq3}
\xpom \equiv \frac{q \cdot (P - p_{_Y})}{q \cdot P} = 
\frac{\mx^2 + Q^2 - t}{W^2 + Q^2 - m_p^2}
\hspace{1.5cm}
\beta \equiv \frac{Q^2}{2 q \cdot (P - p_{_Y})} =
\frac{Q^2}{\mx^2 + Q^2 - t}  \ ,
\end{eqnarray}
where $m_p$ is the proton mass.
Here, $\xpom$ can be interpreted as the fraction of the proton beam
momentum transferred to the system $X$
and $\beta$ may be considered as the 
fraction of the exchanged longitudinal momentum that is carried by the
quark coupling to the photon.

Where leading baryons are directly tagged (see section~\ref{techniques}), 
the measured 
quantities are the transverse
momentum $\pt$ and energy $E_p^{\prime}$ of the final state proton or neutron. 
The kinematic variables used to describe the process are
\begin{eqnarray}
\label{eq4}
\xl \equiv 1 - \frac{q . (p - p^{\prime})}{q.p} 
\simeq \frac{E_p^{\prime}}{E_p}
\hspace{1.7cm}
t \equiv (p - p^{\prime})^2 \simeq - \frac{\pt^2}{\xl} - 
(1 - \xl) \left[ \frac{m_N^2}{\xl} - m_p^2 \right] \ ,
\end{eqnarray}
where $p^{\prime}$ is the 4-vector of the final state nucleon and
$m_N$ is its mass. 
Provided the final state nucleon is
exclusively produced at the proton vertex,
the definitions (\ref{eq2}) and (\ref{eq4}) of $t$ are equivalent
and $\xl = 1 - \xpom$.

\section{Experimental Techniques}
\label{techniques}

Much of the data presented here is selected on the basis of
a large rapidity gap adjacent to the outgoing proton beam,
identified by an absence of activity  
in the more forward~\footnote{In the HERA coordinate system, the `forward' 
positive $z$ direction is that of the outgoing proton beam and corresponds to 
positive values of rapidity.} parts of the detectors.
The resulting data samples are
dominated by the case where $Y$ is a proton. 
This approach yields high acceptance over a wide kinematic region.
Since the
size of the rapidity gap separating the systems $X$ and $Y$ decreases as 
$\mx$ grows, measurements by this method are restricted
to the region $\xpom \lapprox 0.05$ in which diffraction is expected to be
dominant.

In a second experimental method, final state baryons are  
detected directly. Purpose-built forward proton
spectrometers (FPSs) are exploited to detect and measure final state 
protons \cite{zeus:fps1,H1:LB}. 
Calorimetry is also installed along the forward beam-line
to detect and measure leading neutrons \cite{zeus:fnc,H1:LB}.
The direct tagging method allows the system $Y$ to be positively identified 
as a proton or neutron and
is the only way of measuring the $t$ distribution
for the photon dissociation process. It also allows
access to an enhanced region of $\xpom$. However, the incomplete
acceptances of the forward proton and neutron detectors limit the
available statistics.

The exclusive production of vector mesons is 
identified through the absence
of any activity in the 
central and backward parts of the detectors beyond that associated with the 
vector meson decay. 
The charged decay products of the vector mesons are detected with high 
resolution in central tracking devices. For the measurements described here,
the decay channels $\rho^0 \rightarrow 
\pi^+ \pi^-$, $\phi \rightarrow K^+ K^-$, $J/\psi \rightarrow \mu^+ \mu^-$
and $J/\psi \rightarrow e^+ e^-$ are studied.  
Measurements have also been made using the decays $\omega \rightarrow 
\pi^+ \pi^- \pi^0$ \cite{omega}, 
$\rho^{\prime} \rightarrow \pi^+ \pi^- \pi^+ \pi^-$ \cite{rho:prime} and
various decay modes of $\psi (2S)$ \cite{psi:prime}. Observations of
the $\Upsilon$ have also recently been reported \cite{upsilon}.  

In the study of photon dissociation by the rapidity gap or leading baryon
tagging methods, 
the system $X$ is measured in the central parts of the detectors after 
requiring that there be no calorimetric activity forward of a given 
pseudorapidity.
A third selection procedure is favoured by ZEUS for studies of photon
dissociation \cite{zeus:mx}. Without explicitly requiring a forward rapidity
gap, the diffractive signal is extracted from the low mass tail in the
distribution in hadronic
invariant mass $\mx$ visible in the central components of the 
detector (see section~\ref{ddis}).


\section{Dependences on {\boldmath $W$} and {\boldmath $t$}}

\subsection{Parameterising Colour Singlet Exchange Cross Sections}

The language of Regge phenomenology is generally used to discuss the kinematic
dependences of colour singlet exchange cross sections at HERA. This does not
necessarily imply the exchange of universal Regge poles.  
It does, however,
provide a convenient means of parameterising the data and comparing
processes involving the scattering of different
particles at different values of $t$ or $Q^2$.

Usually, only the minimal asymptotic Regge assumption is 
needed,\footnote{Triple Regge
analysis has been applied to diffractive dissociation in
photoproduction \cite{mx:gammap,zeus:mx}.} which states that
with $t$, $\mx$ and $Q^2$ fixed,
\begin{eqnarray}
\frac{{\rm d} \sigma}{{\rm d} t {\rm d} \mx^2} \propto 
\left( \frac{1}{\xpom} \right)^{2 \alphapom(t) - 2} \ ,
\label{minimal}
\end{eqnarray}
where $\alphapom(t)$ 
is the {\em effective leading Regge trajectory}
describing the process. Varying $\xpom$ at fixed $\mx$, $Q^2$ and $t$ is
equivalent to varying $W^2$ (equation~\ref{eq3}). Since $W^2 \gg Q^2$
throughout most of the HERA kinematic domain, $1 / \xpom$ in 
equation~\ref{minimal} is often replaced simply by $W^2$.

In order to additionally model the $t$ dependence, it is necessary to introduce
the form factors of the interacting particles. 
These are usually
approximated to empirically motivated 
exponential functions, such that at fixed $\mx$ and $Q^2$,
the full $W$ and $t$ dependence is given by
\begin{eqnarray}
\frac{{\rm d} \sigma}{{\rm d} t {\rm d} \mx^2} \propto 
\left( \frac{1}{\xpom} \right)^{2 \alphapom(t) - 2} \ e^{b_0 t} \ \sim
\left( W^2 \right)^{2 \alphapom(t) - 2} \ e^{b_0 t} \ .
\label{vmregge}
\end{eqnarray}

For soft diffractive processes,
the leading trajectory 
takes the universal linear form 
$\alphapom(t) \sim \alpha(0) + \alpha^{\prime} t \sim 
1.08 + 0.25 \ t$ \cite{soft}. Where hard scales are
present, the effective pomeron intercept $\alphapom(0)$ is 
expected to increase and the trajectory slope $\alpha^{\prime}$ 
is expected to decrease \cite{vm:gg}.

\subsection{Vector Meson Energy and Momentum Transfer Dependences}
\label{softhard}

\begin{figure}[htb] \unitlength 1mm
 \begin{center}
   \begin{picture}(160,100)
     \put(-5,0){\epsfig{file=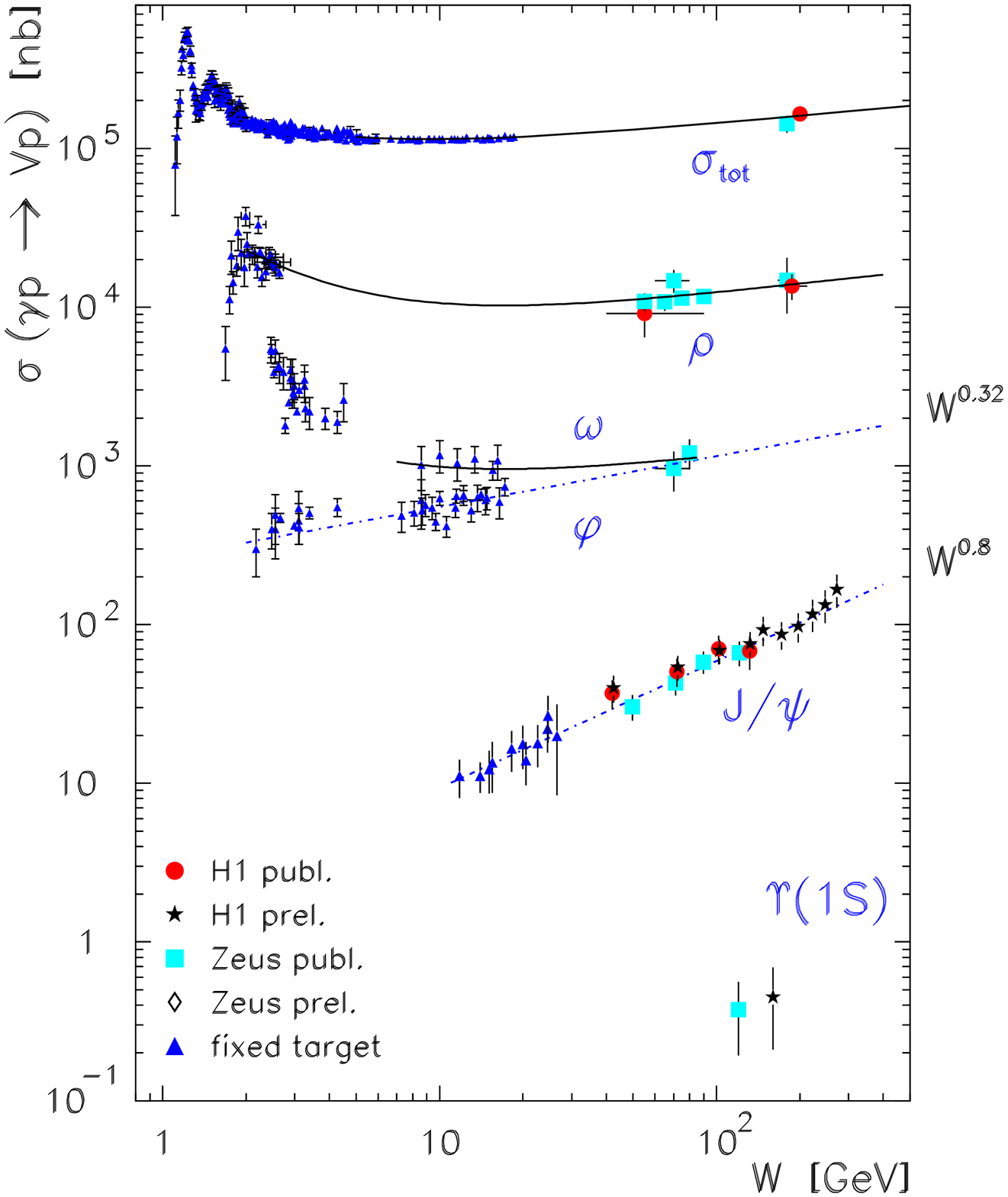,width=0.6\textwidth}}
     \put(82,55){\epsfig{file=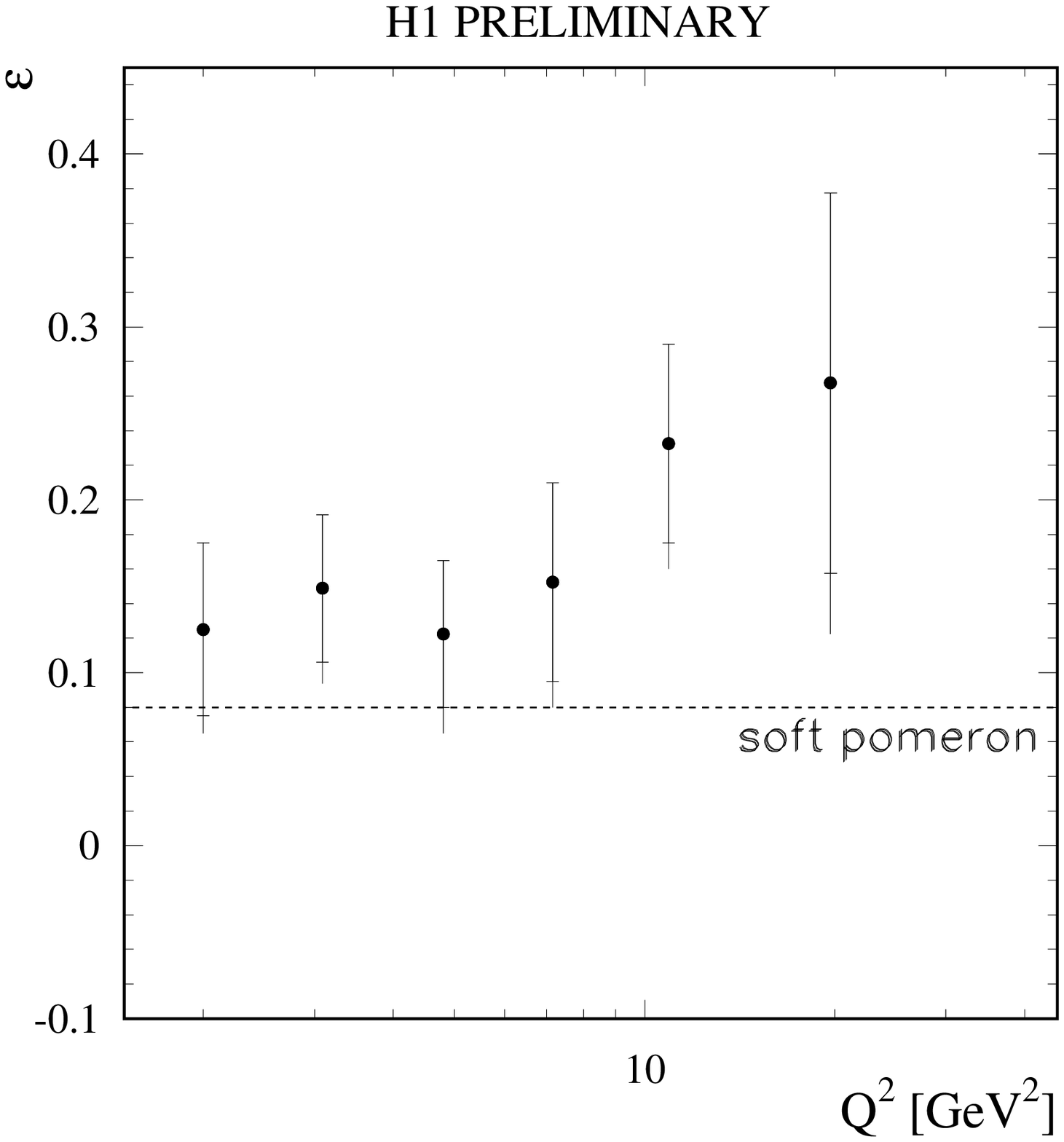,width=0.35\textwidth}}
     \put(80,0){\epsfig{file=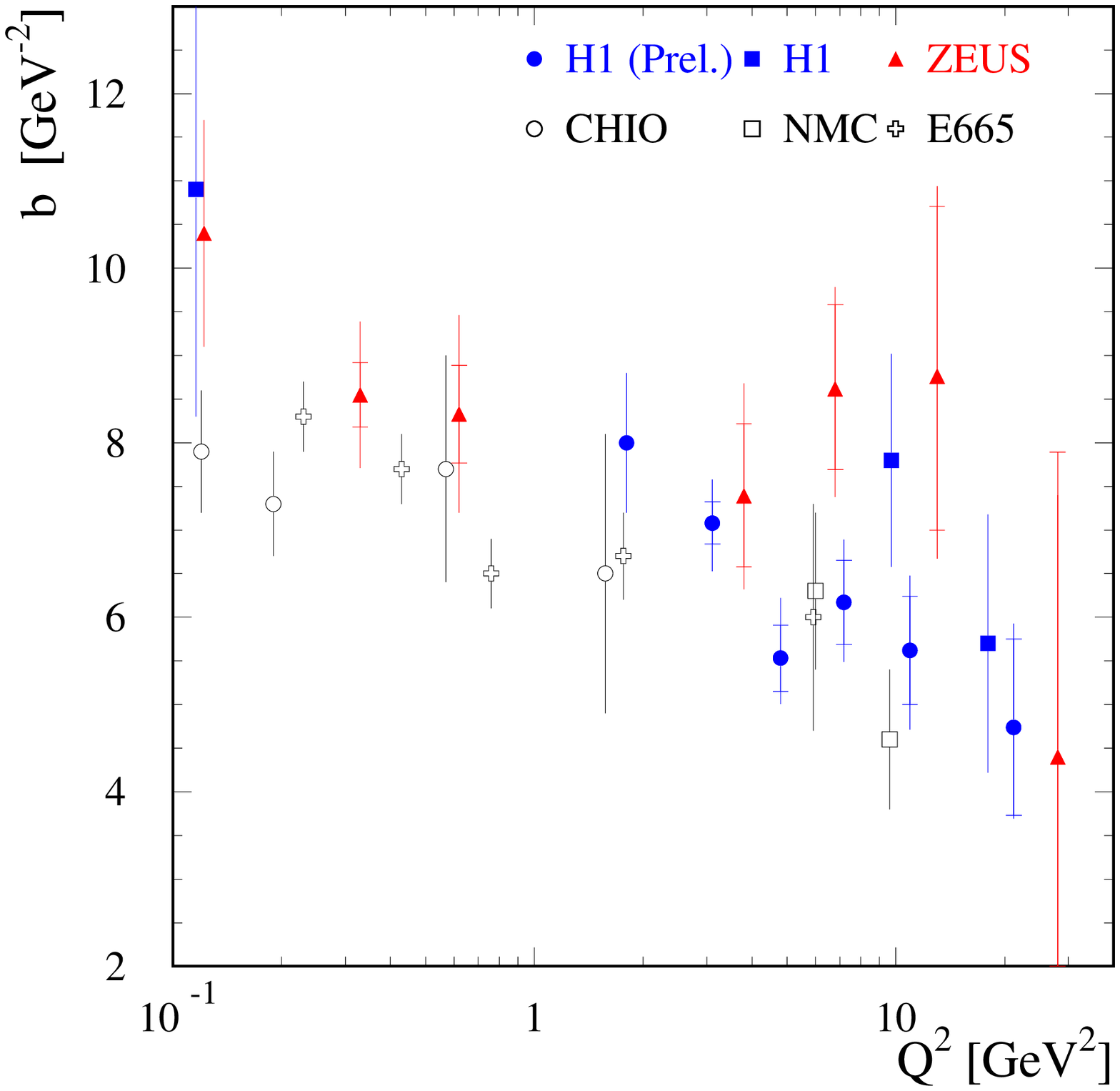,width=0.41\textwidth}}
     \put(15,90){\large{\bf{(a)}}}
     \put(90,95){\large{\bf{(b)}}}
     \put(90,15){\large{\bf{(c)}}}
   \end{picture}
 \end{center}
  \caption  {(a) Various vector meson cross sections shown as a function of
$W$ at $Q^2 = 0$, together with the total photoproduction cross section.
(b) The $Q^2$ dependence of the parameter $\epsilon = \alphapom(0) - 1$,
extracted from fits to the $W$ dependence of 
exclusive $\rho^0$ electroproduction. (c) The $Q^2$ dependence of the slope
parameter $b$ for exclusive $\rho^0$ electroproduction.}
  \label{vmxsecs}
\end{figure}

Figure~\ref{vmxsecs}a shows the centre of mass energy dependence of various
$t$-integrated vector meson cross sections at 
$Q^2 = 0$ \cite{rho:gammap,zeus:gprho,omega,zeus:phi,h1:jpsipub,zeus:jpsi,h1:jpsi,upsilon}, after correcting for the 
small proton dissociation contributions. In common with the total 
cross section at $Q^2 = 0$ \cite{stot}, the energy
dependence of the $\rho^0$ photoproduction cross section follows 
that expected
for the exchange of the universal soft pomeron of hadronic physics. In contrast
to this, for $J/\psi$ photoproduction, where a hard scale corresponding to
the charm quark mass is introduced, the $W$ dependence of the cross section 
is much steeper.

Figure~\ref{vmxsecs}b shows the effective pomeron intercept as a function of 
$Q^2$ for the process $\gamma^* p \rightarrow \rho^0 p$, as extracted from 
fits to equation~\ref{vmregge} after integration over $t$. 
There is evidence for a steepening of the $W$ depedendence of
the rho electroproduction cross section as $Q^2$ increases.
It is thus apparent that the introduction of hard scales such as 
$Q^2$ or a heavy quark mass leads to an increase in the effective 
$\alphapom(0)$ describing the vector meson production energy dependence.

The $t$ dependence of vector meson production is studied by fitting
data to the form ${\rm d} \sigma / {\rm d} t \propto e^{b t}$, where in the
context of equation~\ref{vmregge}, $b = b_0 + 2 \alpha^{\prime} 
\ln (W^2 / {\rm GeV^2})$. 
The extracted value of $b$ for $\rho$ production is shown 
as a function of $Q^2$ in figure~\ref{vmxsecs}c. There is a clear decrease
of $b$ with increased $Q^2$. This indicates that the 
$\gamma^* \rightarrow V$ transition becomes an increasingly short distance
process as $Q^2$ is increased.
For the case
of $J/\psi$ photoproduction, $b \sim 4.5 \ {\rm GeV^{-2}}$, similar to that
for the $\rho$ at large $Q^2$, again indicating that exclusive
$J/\psi$ electroproduction
is already a hard process at $Q^2 = 0$. 


\subsection{The Diffractive Dissociation Cross Section at Low 
{\boldmath $\xpom$}}
\label{ddis}

\begin{figure}[htb] \unitlength 1mm
 \begin{center}
   \begin{picture}(160,65)
     \put(2,60){\epsfig{file=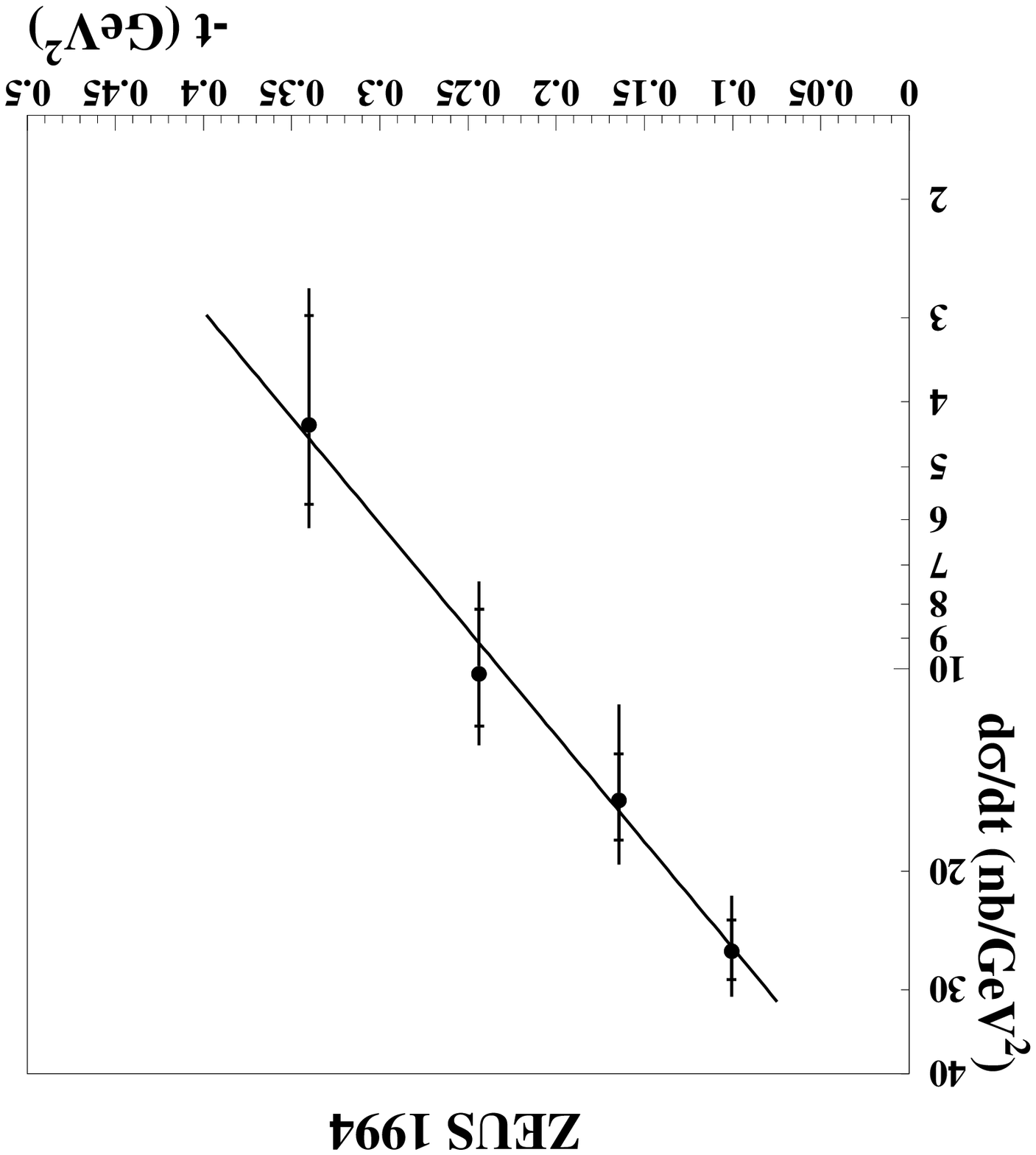,angle=180,width=0.037\textwidth}}
     \put(60,35){\epsfig{file=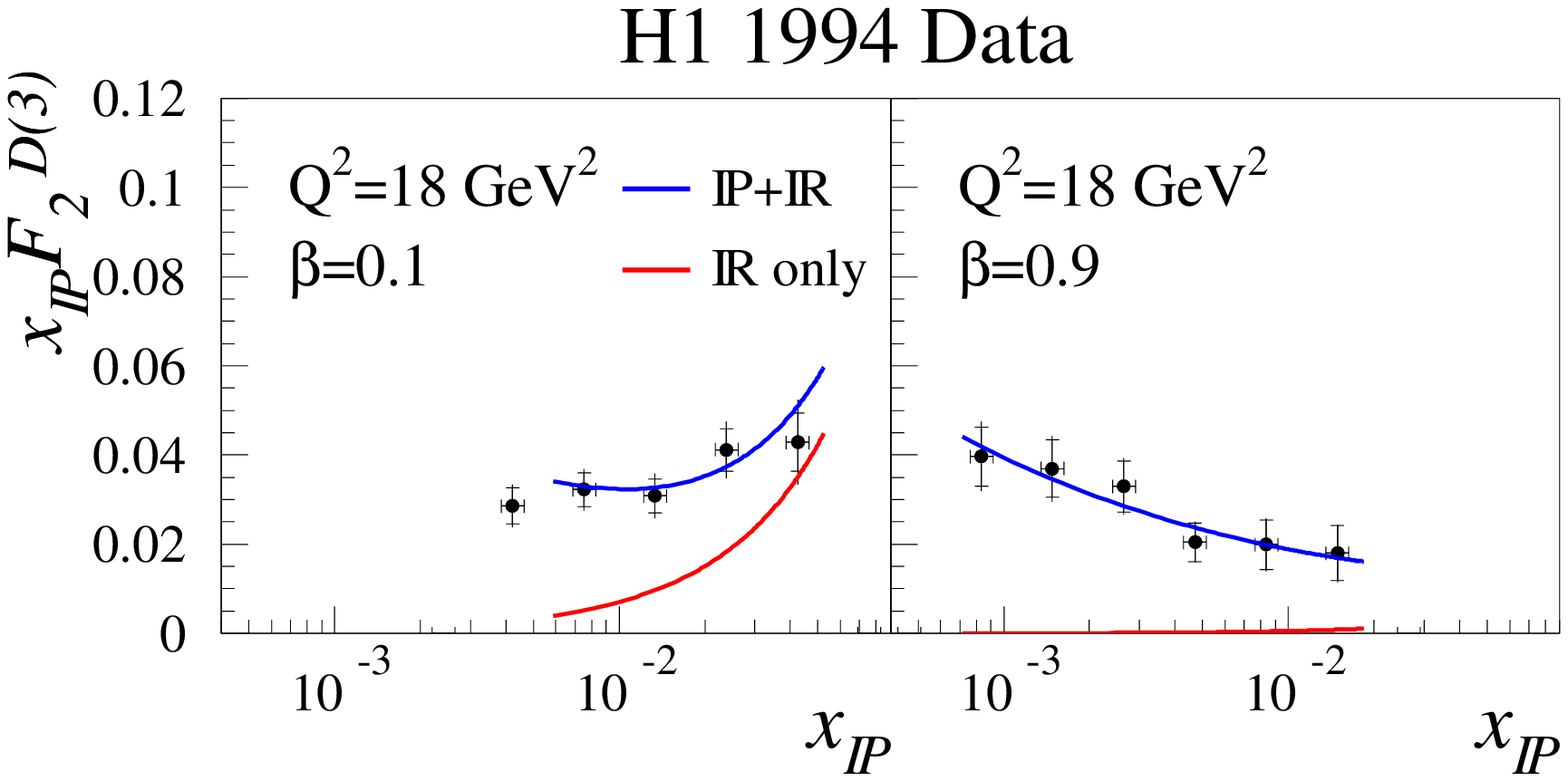,width=0.6\textwidth}}
     \put(55,-15){\epsfig{file=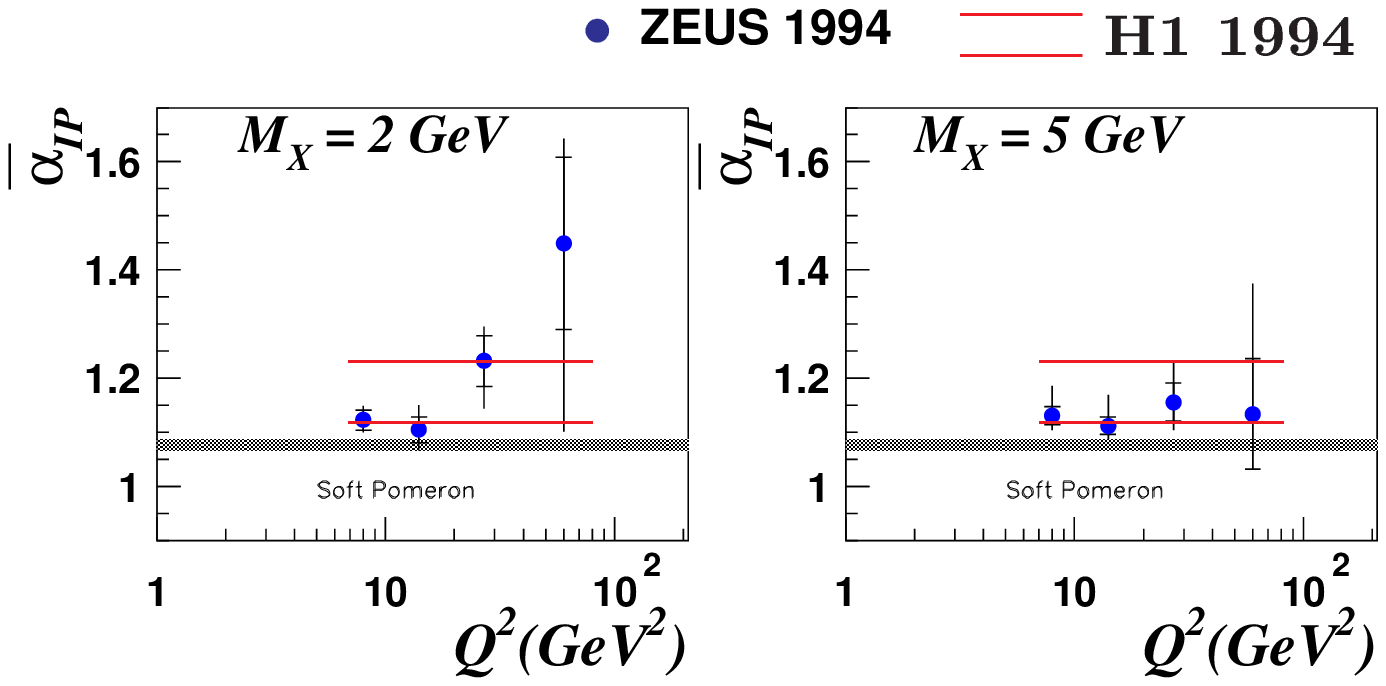,width=0.65\textwidth}}
     \put(15,15){\large{\bf{(a)}}}
     \put(123,56){\large{\bf{(b)}}}
     \put(123,20){\large{\bf{(c)}}}
   \end{picture}
 \end{center}
  \caption  {(a) Differential $t$-distribution in the
kinematic region $5 < Q^2 < 20 \ {\rm GeV^2}$, $0.015 < \beta < 0.5$ and
$\xpom < 0.03$, obtained using the ZEUS FPS. (b) Example H1 measurements of 
$\xpom \cdot F_2^{D(3)}$, together
with the results of a fit to the parameterisation of equations~\ref{fluxparam}
and~\ref{f2dparam}. (c) Comparison of H1 and ZEUS extractions of 
$\alphapom(0)$ for virtual photon dissociation, 
together with the value expected for the `soft' pomeron that 
governs hadronic interactions in the absence of hard scales.}
  \label{f2d4}
\end{figure}

The $t$ dependence of the virtual photon dissociation process has been 
measured in the diffraction dominated low $\xpom$ region using the ZEUS 
FPS \cite{zeus:fps}. The data are shown in 
figure~\ref{f2d4}a, together with a fit to the form 
${\rm d} \sigma / {\rm d} t \propto e^{b t}$. The result is
$b = 7.2 \ \pm \ 1.1 \ ({\rm stat.}) \ ^{+0.7}_{-0.9} \ ({\rm syst.})
\ {\rm GeV^{-2}}$, revealing a highly peripheral scattering characteristic
of a diffractive process.

Virtual photon dissociation data obtained by the rapidity gap technique are
usually presented in the form of
a three dimensional structure function 
$F_2^{D(3)}(\beta, Q^2, \xpom)$, defined in close
analogy to the inclusive proton structure function as
\begin{eqnarray}
  F_2^{D(3)} (\beta,Q^2,\xpom) = 
  \frac{\beta Q^4}{4 \pi \alpha^2} \ \frac{1}{(1 - y + y^2 / 2)} \
  \int_{t_0}^{t_{\rm min}} {\rm d} t \
  \frac{{\rm d}^4 \sigma^{ep \rightarrow eXY}}{{\rm d} \beta \ 
  {\rm d} Q^2 \ {\rm d} \xpom {\rm d} t} \ .
\end{eqnarray}
Measurements of $F_2^{D(3)}$
have been made by H1 using the rapidity gap technique
at 356 points in $\beta$, $Q^2$ 
and $\xpom$ space, spanning the region $0.001 < \beta < 0.9$ and 
$0.4 < Q^2 < 800 \ {\rm GeV^2}$,
integrated over $\my < 1.6 \ {\rm GeV}$ with 
$t_0 = - 1.0 \ {\rm GeV^2}$ \cite{H1:F2D3,H1:F2Dprelim}. 
For a single universal exchange, a factorisation of the $\xpom$ dependence
is expected of the form \cite{diff:hardscat}
\begin{eqnarray}
  \label{factorise}
  F_2^{D(3)} \ = \ f_{\pom / {\rm p}} (\xpom) \ F_2^{\pom} (\beta, Q^2) \ ,
\end{eqnarray}
where, for a cross section differential in $\xpom$, after integrating 
equation~\ref{vmregge} over $t$,
\begin{eqnarray}
  f_{\pom / {\rm p}} (\xpom) \propto 
\int_{-1 {\rm GeV^2}}^{t_0 (\xpom)} \left(
\frac{1}{\xpom} \right)^{2 \alphapom(t) - 1} e^{B_{\pom} \, t} \ {\rm d} t 
\label{fluxparam} 
\end{eqnarray}
parameterises the flux factor for the exchange $i$ from the proton. Here,
$B_{\pom} = b_0 + 2 \alpha^{\prime} \ln (1 / \xpom)$ in 
equation~\ref{vmregge}. In equation~\ref{factorise},
$F_2^{\pom}(\beta, Q^2)$ may be interpreted as being proportional to the 
structure function of the pomeron \cite{diff:hardscat}.


When viewed in close detail, the data do no quite obey the factorisation
of equation~\ref{factorise}.
As an example, data at $Q^2 = 18 \ {\rm GeV^2}$ and two 
different values of $\beta$ are shown in 
figure~\ref{f2d4}b. At fixed $Q^2$, a variation in the
$\xpom$ dependence of $F_2^{D(3)}$ 
is observed as $\beta$ changes. 
In a Regge model, this
is interpreted in
terms of the exchange of sub-leading trajectories in addition to
the pomeron when $\xpom$ becomes large. 
Table~\ref{reggeons} summarises the trajectories that have
most commonly been considered in soft hadronic physics.
In a more complete Regge treatment, 
the differential
structure function then follows~\footnote{Interference terms are also
possible, in particular, that between $\pom$ and $f$. In \cite{H1:F2D3}, 
there was
found to be little sensitivity to the presence or absence of this term.}
\begin{eqnarray}
  F_2^{D(3)} \ = \sum_{i=\pom,\reg,\pi \ldots}
f_{i / {\rm p}} (\xpom) \ F_2^{i} (\beta, Q^2)
\label{f2dparam}
\end{eqnarray}
where $f_{i / {\rm p}}$ is a flux factor for the exchange $i$,
similar in form to equation~\ref{fluxparam}.

\begin{footnotesize}
\begin{table}[htb]
\begin{center}
\begin{tabular}{|c||c|c|c|c|} \hline
TRAJECTORY & APPROX. $\alpha(0)$ & APPROX $\alpha^{\prime}$ & 
ISOSPIN & $p$ : $n$ \\ \hline \hline
$\pom$        & 1   & 0.25 & 0 & 1 : 0 \\ \hline
$f$, $\omega$ & 0.5 & 1.0  & 0 & 1 : 0 \\ \hline 
$\rho$, $a$   & 0.5 & 1.0  & 1 & 1 : 2 \\ \hline
$\pi$         & 0   & 1.0  & 1 & 1 : 2 \\ \hline
\end{tabular}
\caption{Summary of commonly discussed Regge trajectories, 
together with their approximate
intercepts and slopes, their isospin and the expected
ratio of fluxes for leading proton and neutron production. The exchange
degenerate $f$, $\omega$, $\rho$ and $a$ trajectories are collectively 
referred to as $\reg$.}
\label{reggeons}
\end{center}
\end{table}
\end{footnotesize}

Allowing for the exchange of one further trajectory ($\reg$) in addition to the
pomeron is sufficient to obtain a good description of $F_2^{D(3)}$ throughout
the measured kinematic range. In
fits to intermediate $Q^2$ data \cite{H1:F2D3} based on 
equation~\ref{f2dparam}
in which $\alphapom(0)$, $\alphareg(0)$ and the values of
$F_2^{\pom}$ and $F_2^{\reg}$ at each ($\beta$, $Q^2$) 
point are free parameters,
the intercept of the secondary trajectory is found to be
$\alphareg(0) = 0.50 \pm 0.11 \ ({\rm stat.}) \pm 0.11 \ ({\rm syst.})
^{+ 0.09}_{-0.10} \ ({\rm model})$, consistent with the exchange of the $f$ 
or one of its exchange degenerate partners. 

In their analysis of virtual photon dissociation, ZEUS fit the hadronic 
invariant mass distribution at
fixed $W$ and $Q^2$, to the form 
\begin{eqnarray}
\label{zeuscrap}
{\rm d} {\cal N}  / {\rm d} \ln \mx^2 = D + 
c \, e^{b \ln \mx^2} \ ,
\end{eqnarray} 
where the second term parameterises the contribution from non-diffractive
processes at large $\mx$. The
operationally defined diffractive contribution $D$ is acceptance corrected 
to give the cross section in intervals
of $\mx$, $W^2$ and $Q^2$ for the process
$ep \rightarrow eXY$ with $\my < 5.5 \ {\rm GeV}$ \cite{zeus:mx}. The $W$
dependence of the diffractive contribution at fixed $Q^2$ and
$\mx$~\footnote{The variable $\beta$ is thus also fixed.}
is then used to extract the 
$t$ averaged value of the pomeron intercept, $\overline{\alphapom}$, 
by fitting to the form 
\begin{eqnarray}
\frac{{\rm d} \sigma^{ep \rightarrow e X Y}}{{\rm d} \mx} (\mx,W,Q^2)
\propto \left(W^2 \right)^{2 \ \overline{\alphapom} -2} \ ,
\end{eqnarray}
derived from equation~\ref{minimal}.
The results are shown in figure~\ref{f2d4}c and are compared
with the H1 result, 
$\alphapom(0) = 1.203 \pm 0.020 \ {\rm (stat.)} 
\pm 0.013 \ {\rm (syst.)} \ ^{+ 0.030}_{-0.035} {\rm (model)}$, 
amended to
$\overline{\alphapom}$ assuming 
$b = 7.2 \pm 1.4 \ {\rm GeV^{-2}}$ \cite{zeus:fps}
and $\alphapom^{\prime} = 0.26 \pm 0.26 \ {\rm GeV^{-2}}$.
The two experiments
are found to be in good agreement. The effective pomeron intercept is 
significantly larger than the values $\alphapom \sim 1.1$ obtained from soft
diffractive processes \cite{soft} and is similar to that describing 
exclusive $J/\psi$ photoproduction. To date, there is no evidence for a
variation of the effective $\alphapom(0)$ with $Q^2$
within the deep-inelastic regime.
However,
the values extracted from photon dissociation at $Q^2 = 0$ are
significantly smaller \cite{mx:gammap,zeus:mx}. 

\subsection{Leading baryons at Large {\boldmath $\xpom$}}
\label{lb}

The extensive coverage in final state baryon energy of the leading proton
and neutron detectors allows the study of colour singlet exchange processes
to be extended to the region of comparatively large $\xpom$ (equivalently
small $z$ - see equation~\ref{eq4}), which in Regge models is expected to be
dominated by sub-leading exchanges. 

A large fraction of DIS events are found to contain forward baryons 
in the final state outside the large $\xl$ diffractive region. For example, 
ZEUS find \cite{ZEUS:LB} that 
approximately 12.5 \% of events have a leading proton or neutron with
$0.6 < \xl < 0.9$ scattered through a polar angle $\theta < 0.8 \ {\rm mrad}$.
Fragmentation models \cite{ariadne,lepto}
that are successful in describing energy flow
in the photon fragmentation and central plateau regions fail to 
reproduce the
rates and energy distributions of the low $\xl$ leading baryons.

\begin{figure}[htb] \unitlength 1mm
 \begin{center}
   \begin{picture}(120,48)
     \put(-5,-5){\epsfig{file=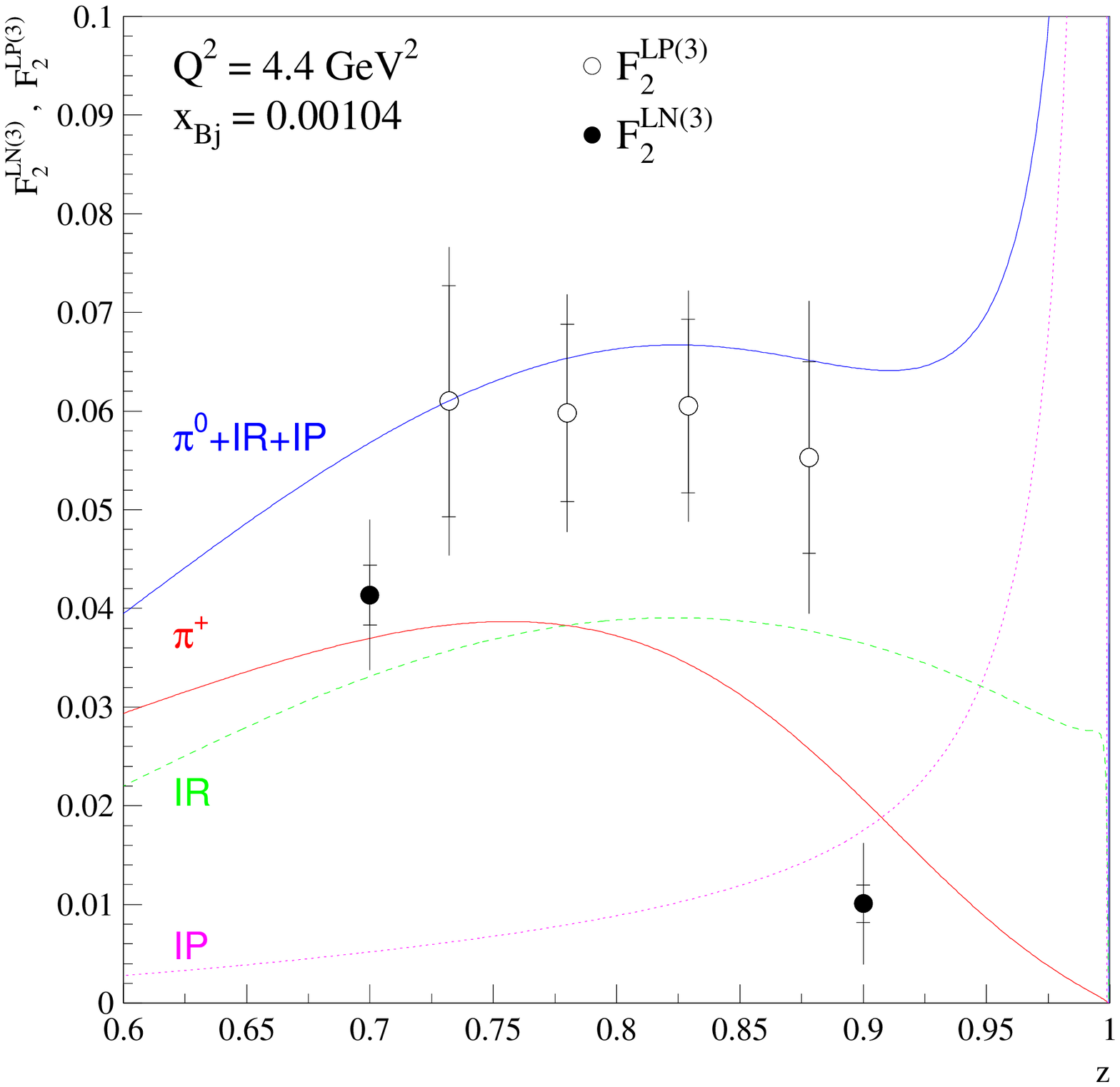,width=0.42\textwidth}}
     \put(47,-42){\epsfig{file=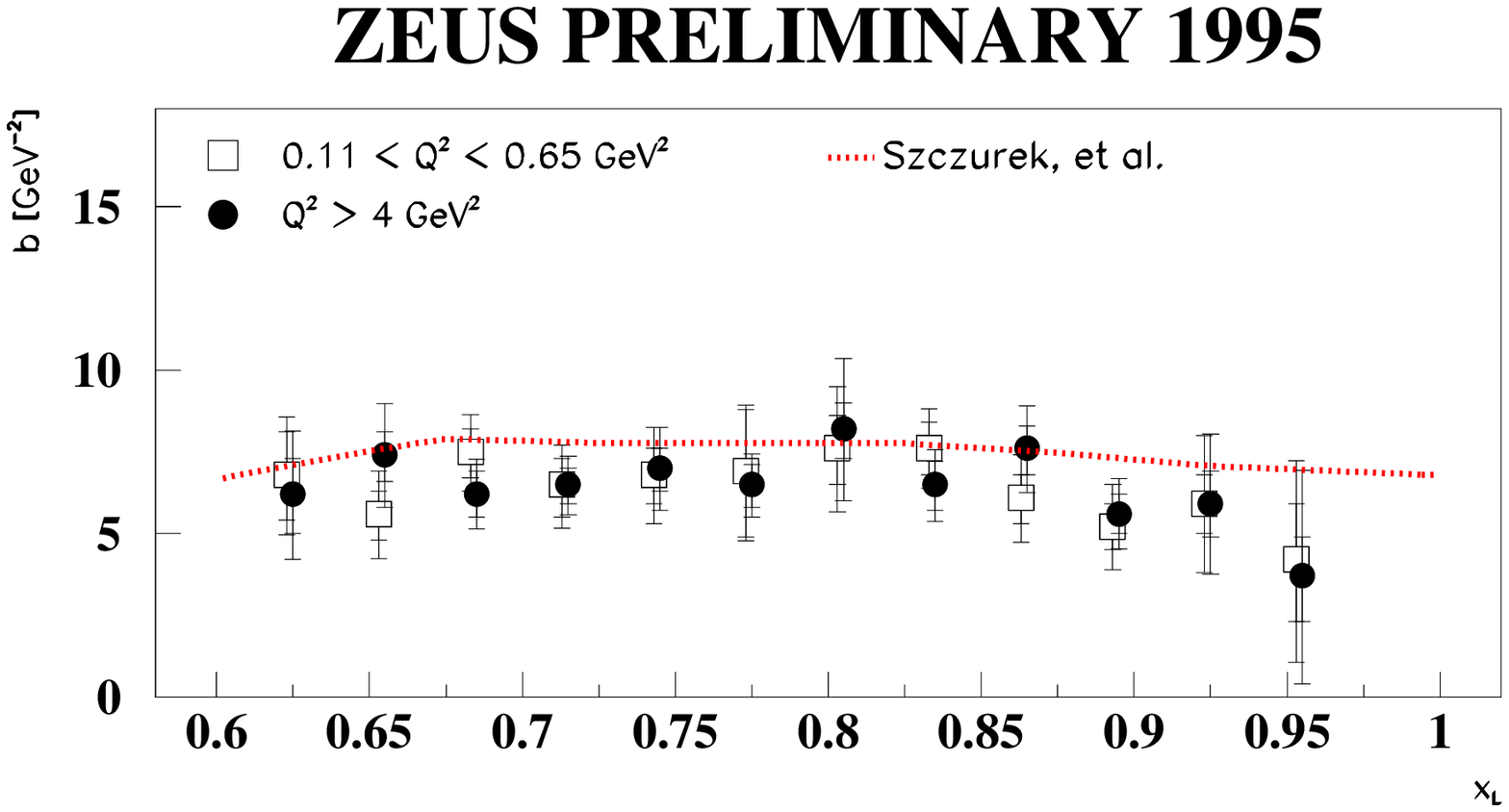,width=0.65\textwidth}}
     \put(13,48){\bf H1 DATA}
     \put(6,48){\bf (a)}
     \put(57.5,38.5){\bf (b)}
   \end{picture}
 \end{center}
  \caption  {(a) Dependence of the leading proton (open points) and 
neutron (closed points)
structure functions on $\xl$ in an example bin of $x$ and $Q^2$.
The decomposition of the data according to a Regge model is
superimposed. (b) Dependence of the $t$ slope parameter
on $\xl$, with a comparison to a similar model.}
  \label{lbregge}
\end{figure}

H1 define leading proton and neutron structure functions, $F_2^{LP(3)}$ 
and $F_2^{LN(3)}$ respectively,
in a similar way to $F_2^{D(3)}$;
\begin{eqnarray}
  F_2^{LB(3)} (x,Q^2,\xl) = 
  \frac{x Q^4}{4 \pi \alpha^2} \ 
  \frac{1}{1 - y + y^2/2} \
  \int_{t_{200}}^{t_{\rm min}} 
  \frac{{\rm d}^4 \sigma^{ep \rightarrow eXN}}{{\rm d} x \ 
{\rm d} Q^2 \ {\rm d} \xl {\rm d} t} \ ,
\end{eqnarray}
where the integration limit $t_{200}$ is defined through equation~\ref{eq4} by
the condition $\pt = 200 \ {\rm MeV}$. The differential 
structure function is measured in the region $0.6 < \xl < 0.9$ in 12 bins
of $x$ and $Q^2$, one of which is shown in 
figure~\ref{lbregge}a \cite{H1:LB}. ZEUS perform 
a similar analysis for data integrated over $x$ in two regions of 
$Q^2$. They have also 
measured the $t$ dependence of leading baryon production at
low $\xl$ \cite{ZEUS:LB}.
Figure~\ref{lbregge}b shows the $\xl$ dependence of the slope parameter $b$,
obtained by fitting the leading proton data
to the standard exponential form 
${\rm d} \sigma / {\rm d} t \propto e^{bt}$. 

Both collaborations compare their low $\xl$ leading baryon 
measurements with simple Regge pole models. 
The trajectories considered are summarised in 
table~\ref{reggeons}. 
The exchange of isoscalar trajectories leads to the
production of leading protons only. Isovector exchanges yield protons and
neutrons in the ratio $1:2$.
Contributions from proton 
dissociative processes are 
neglected, but have been shown to be small \cite{H1:LB}.

The rate of leading proton 
production is larger than that for leading 
neutron production throughout the region 
$0.6 < \xl < 0.9$ (figure~\ref{lbregge}a). This implies that
leading protons in this region are dominantly produced by an isospin-0 
exchange. The contributions from $\rho$ and $a$
exchanges must therefore be small and the models accordingly consider only 
$\pom$, $\reg \equiv f$, $\omega$ and $\pi$ exchange.
With this combination of exchange trajectories, the data are 
compared to models similar in form to equation~\ref{f2dparam}, with
assumptions regarding
flux factors and structure functions as specified
in \cite{lb:regge}. The breakdown of the cross section into different exchange
contributions is shown in figure~\ref{lbregge}a. The predicted $\xl$ dependence
of the $t$ slope parameter is superimposed in figure~\ref{lbregge}b.
Given that there are no free parameters in these models,
the agreement 
with data is remarkably good. For leading proton production, the $\reg$ 
contribution is found to be approximately twice as large as the $\pi$
contribution, with the $\pom$ exchange contribution smaller still for
$\xl \lapprox 0.9$. The neutron production cross section is saturated by the
prediction for $\pi$ exchange. 


\section{Vector Meson Helicity Analyses}
\label{sdme}

Full extractions of the 15 spin density matrix elements governing vector
meson electroproduction \cite{angular:distn}
have recently been made. 
The spin density matrix elements describe the transitions between the
helicity components of the initial state photon and proton and the final
state vector meson and proton. 

\begin{figure}[htb] \unitlength 1mm
 \begin{center}
   \begin{picture}(160,62)
     \put(-5,-20){\epsfig{file=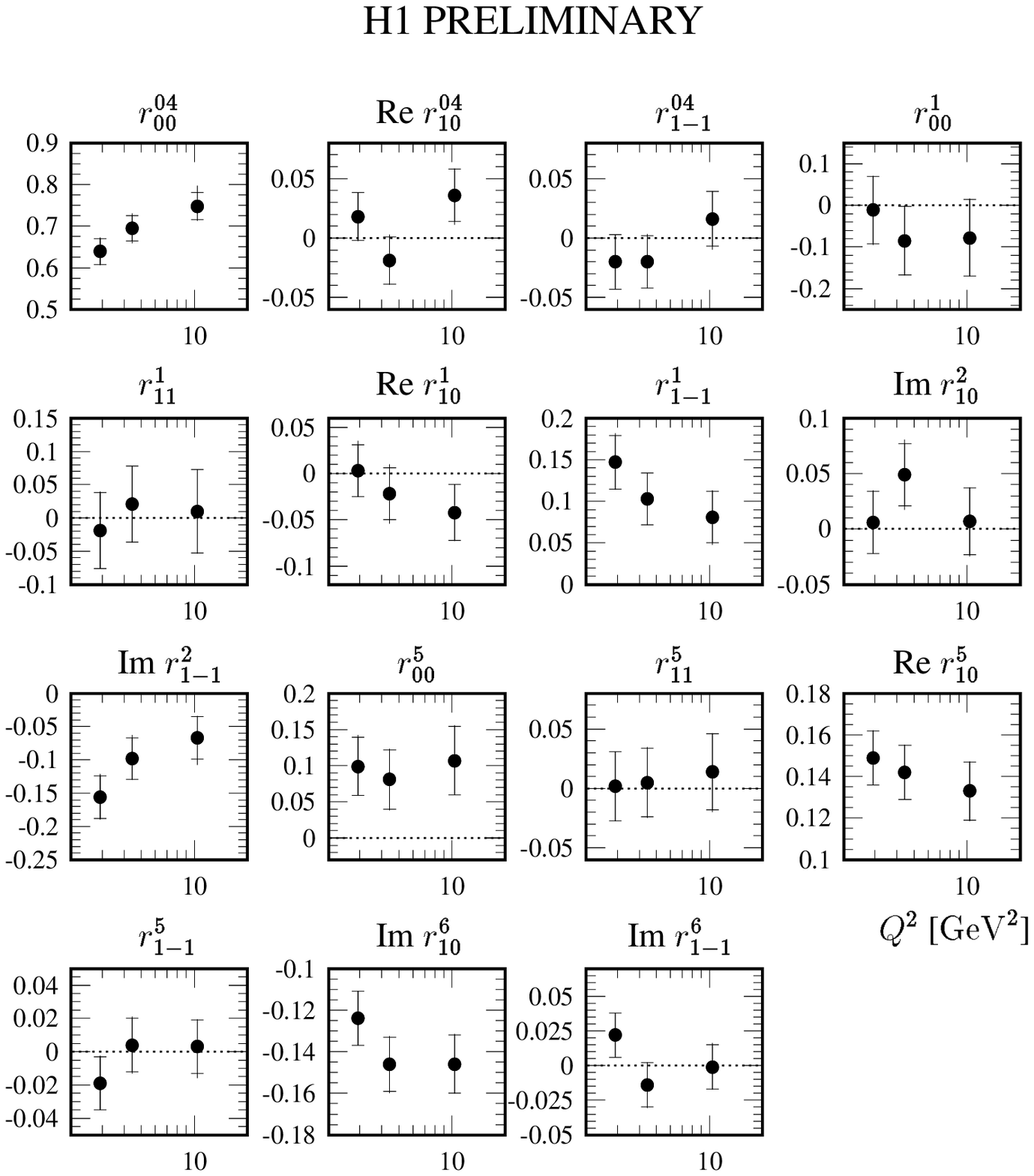,width=0.52\textwidth}}
     \put(75,0){\epsfig{file=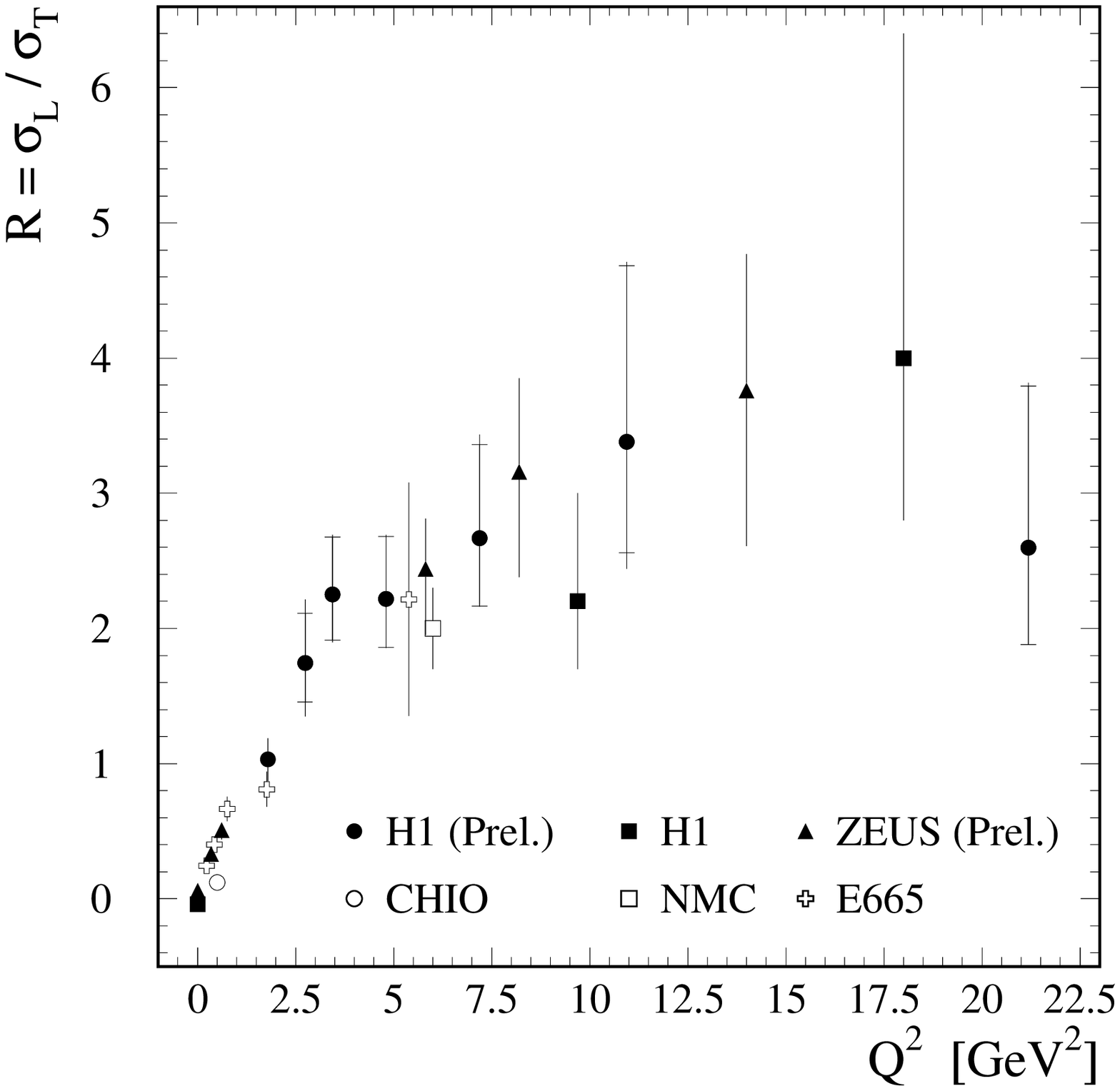,width=0.43\textwidth}}
     \put(10,63){\large{\bf{(a)}}}
     \put(90,63){\large{\bf{(b)}}}
   \end{picture}
 \end{center}
  \caption  {(a) The full set of spin density matrix elements, shown for
$\rho$ electroproduction as a function of $Q^2$. The dashed lines show the
expected values assuming SCHC and NPE. (b) The ratio of longitudinal to
transverse photon induced cross sections $\gamma^* p \rightarrow \rho p$, 
derived from equation~\ref{reqn}, shown as a function
of $Q^2$.}
  \label{helicity}
\end{figure}

Figure~\ref{helicity} shows the 
spin density matrix elements
for the $\rho$ at three different values of $Q^2$ as extracted by
H1 \cite{h1:rho}. The hypothesis of
$s$-channel helicity conservation (SCHC) states that the produced vector
meson should retain the polarisation of the incoming photon. 
The hypothesis of natural parity exchange (NPE) states that
the quantum numbers exchanged in the $t$ channel should have positive parity.
Both hypotheses are usually expected to hold for diffractive 
processes \cite{SCHC}. Where appropriate
in figure~\ref{helicity},
the dashed lines show the expected values of the spin density matrix
elements for the case of SCHC and NPE.
There is a significant deviation from the expected value of zero for the
matrix element $r_{00}^5$. This deviation is also observed by ZEUS for both
$\rho$ and $\phi$ production \cite{zeus:angle}. It implies that 
there is a 
non-zero probability for longitudinally polarised photons to yield transversely
polarised $\rho$ mesons. H1 measure the ratio of single helicity flip to 
helicity conserving amplitudes to be $8 \pm 3 \%$.
The magnitude of the helicity flip amplitude has been predicted in a
QCD inspired model \cite{ivanov}, giving
further confidence that perturbative approaches 
become applicable to vector meson production where hard scales are introduced.

Assuming SCHC, the matrix element
$r_{00}^{04}$ is related to the ratio of cross sections $R$ for 
longitudinal to transverse photons according to
\begin{eqnarray}
  R = \frac{\sigma_{_{\rm L}}}{\sigma_{_{\rm T}}} =
\frac{1}{\epsilon} \ \frac{r_{00}^{04}}{1 - r_{00}^{04}} \ ,
  \label{reqn}
\end{eqnarray}
where the polarisation parameter $\epsilon \sim 0.99$ at HERA. The breaking of
s-channel helicity conservation has only a small effect on the value of $R$
extracted by this method. Figure~\ref{helicity}b shows a compilation of
HERA and fixed target results for $R$, extracted using equation~\ref{reqn}, 
as a function of $Q^2$. The increase in
$R$ with $Q^2$
is found to flatten at large $Q^2$.


\section{Partonic Interpretations of Colour Singlet Exchange}

\subsection{Vector Mesons and the Gluon Structure of the Proton}

Diffractive vector meson
production
follows a similar description to elastic hadron-hadron scattering where no
hard scales are present. However, when the scales $Q^2$, $t$ or the mass of the
valence quarks in the vector meson become large, deviations from soft
pomeron behaviour become apparent (see sections~\ref{softhard} and~\ref{sdme}).
In such cases, it is natural to attempt to
make a perturbative QCD description. The usual approach is to consider the
diffractive scattering of the $q \bar{q}$ fluctuation of the photon, which
subsequently collapses into the vector meson state.
The simplest way to generate a colour
singlet exchange at the parton level is via the exchange of a pair of gluons.
Several authors have built models of hard vector meson production
that contain the square of the gluon distribution 
of the proton \cite{vm:gg,ivanov}.
These models are able to give good descriptions of the
data \cite{zeus:rho,h1:jpsi}.
Particularly for the $\rho$, the success of two-gluon
exchange models 
is tempered by theoretical uncertainties such as that
associated with the vector meson wavefunction.

\subsection{Two Gluon Exchange Models and Virtual Photon Diffractive 
Dissociation}

\begin{figure}[htb] \unitlength 1mm
 \begin{center}
   \begin{picture}(150,46)
     \put(75,-6){\epsfig{file=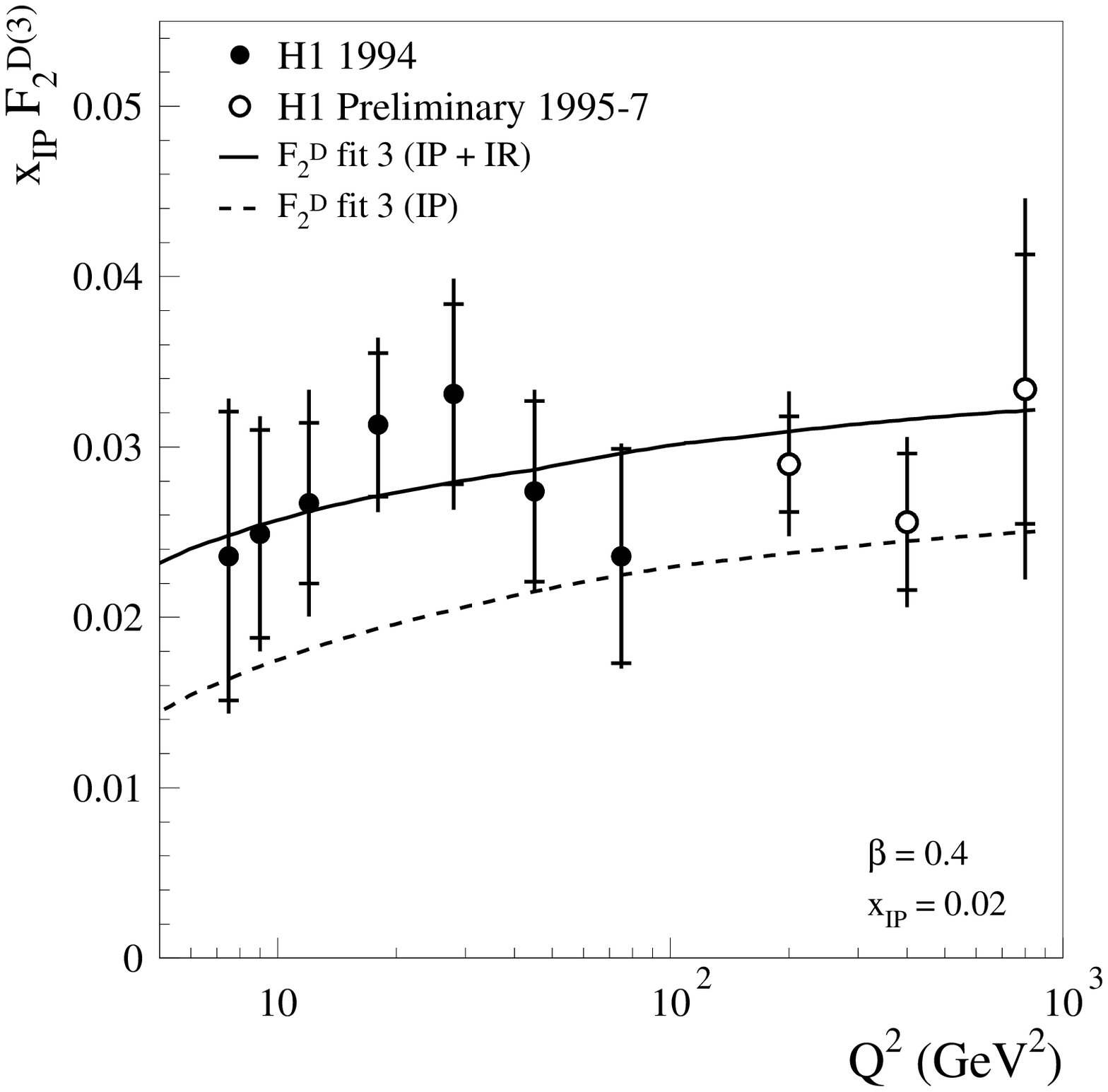,width=0.45\textwidth}}
     \put(0,-6){\epsfig{file=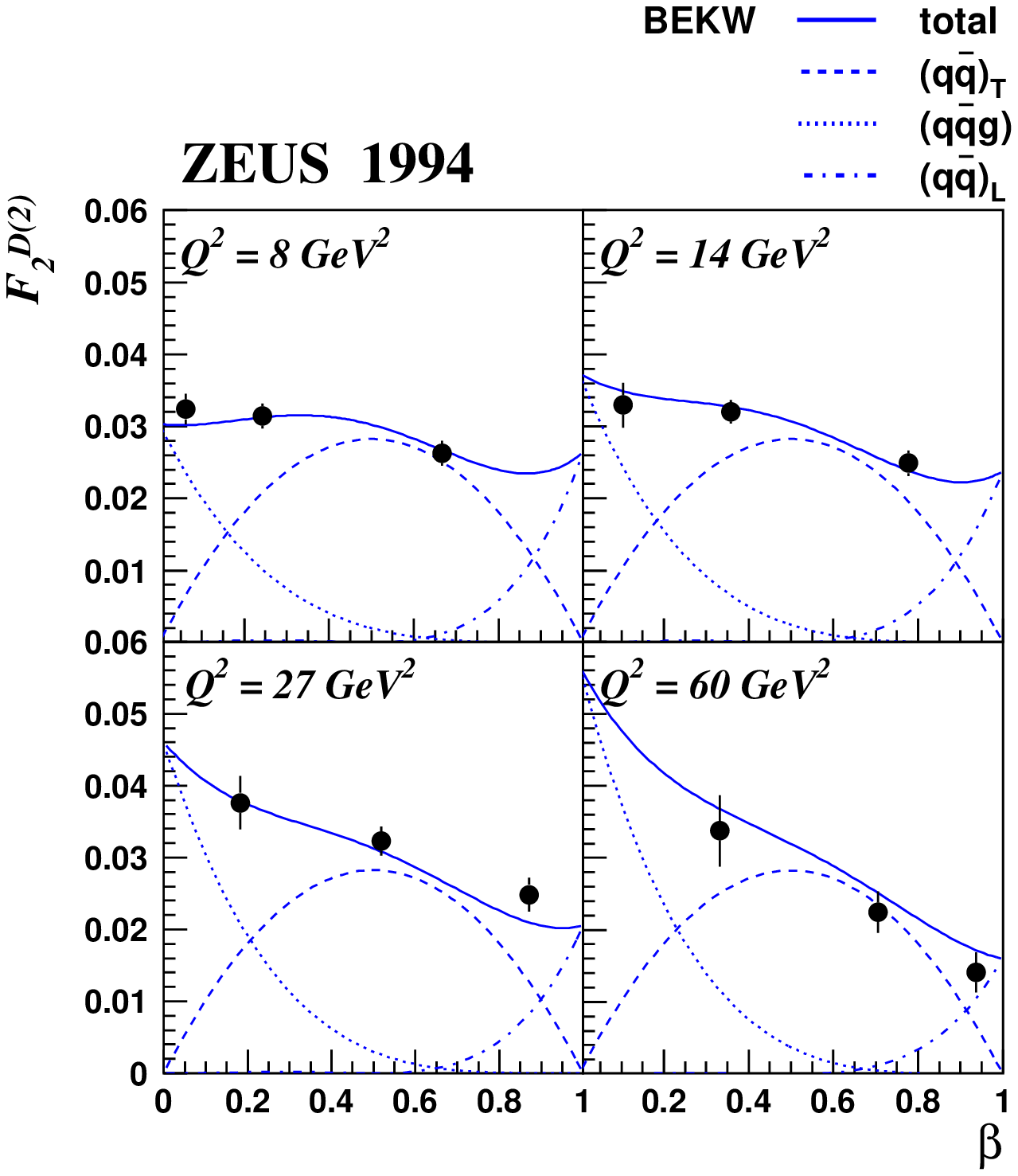,width=0.45\textwidth}}
     \put(10,50){\large{\bf{(a)}}}
     \put(86,5){\large{\bf{(b)}}}
   \end{picture}
 \end{center}
  \caption  {(a) ZEUS measurement of the $\beta$ dependence of $F_2^{D(3)}$
at various values of $Q^2$ and $\xpom = 0.0042$, compared to a model based
on the diffractive scattering of
partonic fluctuations of the photon. (b) H1 measurement of the
$Q^2$ dependence of $\xpom F_2^{D(3)}$ at $\beta = 0.4$ and 
$\xpom = 0.02$. The results of a DGLAP fit to the 1994 data 
in which both quarks and gluons contribute at the starting scale for evolution
are also shown.}
  \label{qcd}
\end{figure}

The success of two-gluon exchange models in describing hard vector meson
production motivates a similar approach to virtual photon diffractive
dissociation. The added complication compared to
the vector meson process is that more complex partonic fluctuations of the
photon such as $q \bar{q} g$ are expected to play a role \cite{qqbarg} in
addition to the simplest $q \bar{q}$ state.
A recent QCD motivated parameterisation of the diffractive structure 
function in terms of the diffractive scattering of $q \bar{q}$ 
and $q \bar{q} g$ states \cite{bartels} is compared to ZEUS measurements
at a fixed small value of $\xpom$ in figure~\ref{qcd}b. In this model,
the medium $\beta$ region is populated dominantly by
$q \bar{q}$ final states originating from transversely polarised photons.
The $q \bar{q} g$ photon
fluctuations are most significant at low $\beta$. Finally, there is a
significant higher twist contribution at large $\beta$ and low $Q^2$,
generated by the interaction of longitudinally polarised photons. 

\subsection{Diffractive Parton Distributions}
\label{pomsf}

In Regge models that consider the exchanged trajectories as distinct 
partonic systems
\cite{diff:hardscat}, the $\beta$ and $Q^2$ dependences
of $F_2^D$ 
at fixed $\xpom$ are sensitive to the parton distributions of the exchanges.
The $\beta$ dependence is relatively
flat (figure~\ref{qcd}a), with significant contributions at large
fractional momenta. An
example of the $Q^2$ dependence at fixed $\beta$ and $\xpom$ is shown in
figure~\ref{qcd}b. Scaling violations
with positive $\partial F_2^D / \partial \log Q^2$ persist
to large values of $\beta \gapprox 0.4$ and extend to large 
$Q^2 \geq 800 \ {\rm GeV^2}$. 
In factorisable partonic pomeron models,
these features indicate the need for a 
significant `hard' (large $\zpom \equiv x_{g/\pom}$) 
gluon contribution to the pomeron 
structure. 

The H1 fits to the $\xpom$ dependence (equation ~\ref{f2dparam}) 
have been extended to 
describe the $\beta$ and $Q^2$ dependence of all data with 
$4.5 \leq Q^2 \leq 75 \ {\rm GeV^2}$ in terms of parton distributions for the
pomeron. Singlet quark and gluon
distributions are parameterised at a starting scale $Q_0^2 = 3 \ {\rm GeV^2}$
and are evolved to larger $Q^2$ using the
DGLAP equations \cite{H1:F2D3}. The 
results from the best fit of this type are shown in figure~\ref{qcd}b. The
cross section in the newly measured region
$200 \leq Q^2 \leq 800 \ {\rm GeV^2}$ \cite{H1:F2Dprelim} 
is also well described
by the extrapolated predictions based on the fits at lower $Q^2$.
Though the precise shape of the parton distributions are rather uncertain
at large $\zpom$, more than 
$80 \ \%$ of the pomeron momentum is carried by 
gluons throughout the $Q^2$ range studied in all such acceptable fits. 

\subsection{The Pion Structure Function}
\label{pisf}

\begin{figure}[htb] \unitlength 1mm
 \begin{center}
   \begin{picture}(120,42)
     \put(20,-8){\epsfig{file=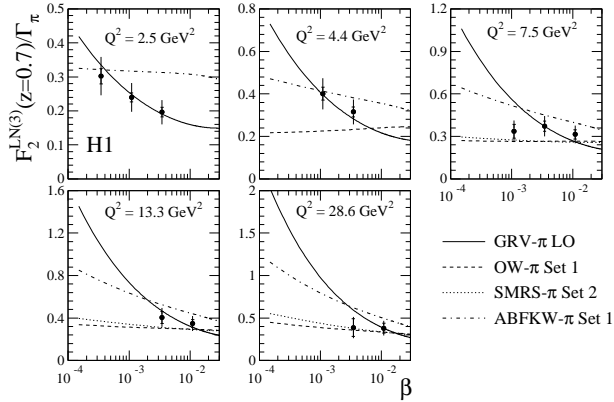,width=0.6\textwidth}}
   \end{picture}
 \end{center}
  \caption  {An extraction of the quantity 
$F_2^{LN(3)} (\beta, Q^2, \xl = 0.7) / \Gamma_\pi (\xl = 0.7)$, which may
be interpreted as the structure function of the pion in the Regge model
described in section~\ref{lb}.}
  \label{f2pi}
\end{figure}

Since the leading neutron cross section can be generated entirely from 
the $\pi$ exchange prediction within the Regge
model described in section~\ref{lb}, the quantity
\begin{eqnarray}
    F_2^{LN(3)} (\beta, Q^2, \xl = 0.7) / \Gamma_\pi (\xl = 0.7)
\end{eqnarray}
may be interpreted as a structure function for the pion, where
$\Gamma_\pi (\xl) = 
\int f_{\pi / p} (\xl, t) \ {\rm d} t$ is the 
pion flux at $z = 0.7$.
Under this assumption, an
extraction of $F_2^{\pi}$ is shown in figure~\ref{f2pi} in a 
previously unexplored low-$x$ region.
The GRV parameterisation \cite{grv:pi} 
matches the data well.

\section{Hadronic Final States of Deep-Inelastic Diffraction}


Many final state observables are sensitive to the partonic structure of the
diffractive interaction and can be used to constrain QCD motivated models. 
The natural frame in which to study the final state is the rest frame of the 
system $X$ ($\gamma^{\star} \pom$ centre of mass frame),
the natural direction in that frame being the $\gamma^{\star} \pom$ 
collision axis.
First HERA results have appeared on
event shapes \cite{thrust}, energy flow and charged particle 
spectra \cite{eflow}, charged particle multiplicities and their 
correlations \cite{multip}, dijet production \cite{zeus:jets,h1:jets} and
charm yields \cite{charm}. 


A summary of most of these measurements can be found in \cite{chicago}.
They have confirmed that there are large
contributions for which the system
$X$ is built from more complex partonic structures than $q \bar{q}$ at lowest
order,
in particular, at large values of $\mx$ and where large transverse momenta 
are generated. In the language of a partonic pomeron, all of these measurements
confirm that the diffractive parton distributions are dominated by gluons 
carrying large fractions of the exchanged momentum, boson-gluon fusion 
(BGF) being
the dominant hard process.

\subsection{Diffractive Dijet Production}

Dijet production is taken here as an example of recent HERA diffractive 
final state data.
The study of dijets is particularly sensitive to the gluon induced 
BGF process. 
The jet transverse momenta $\ptj$ introduce a further hard scale to
the problem, testing the universality of the parton distributions extracted
from $F_2^D$ and allowing hard diffraction to be
tested at $Q^2 = 0$ as well as in DIS. 
Estimators $\xgamj$ and 
$\zpomj$ of the photon and pomeron momenta that are transferred
to the dijet system can be obtained as described in \cite{h1:jets,zeus:jets}. 

\begin{figure}[htbp] \unitlength 1mm
 \begin{center}
   \begin{picture}(120,37)
     \put(39.5,-5){\epsfig{figure=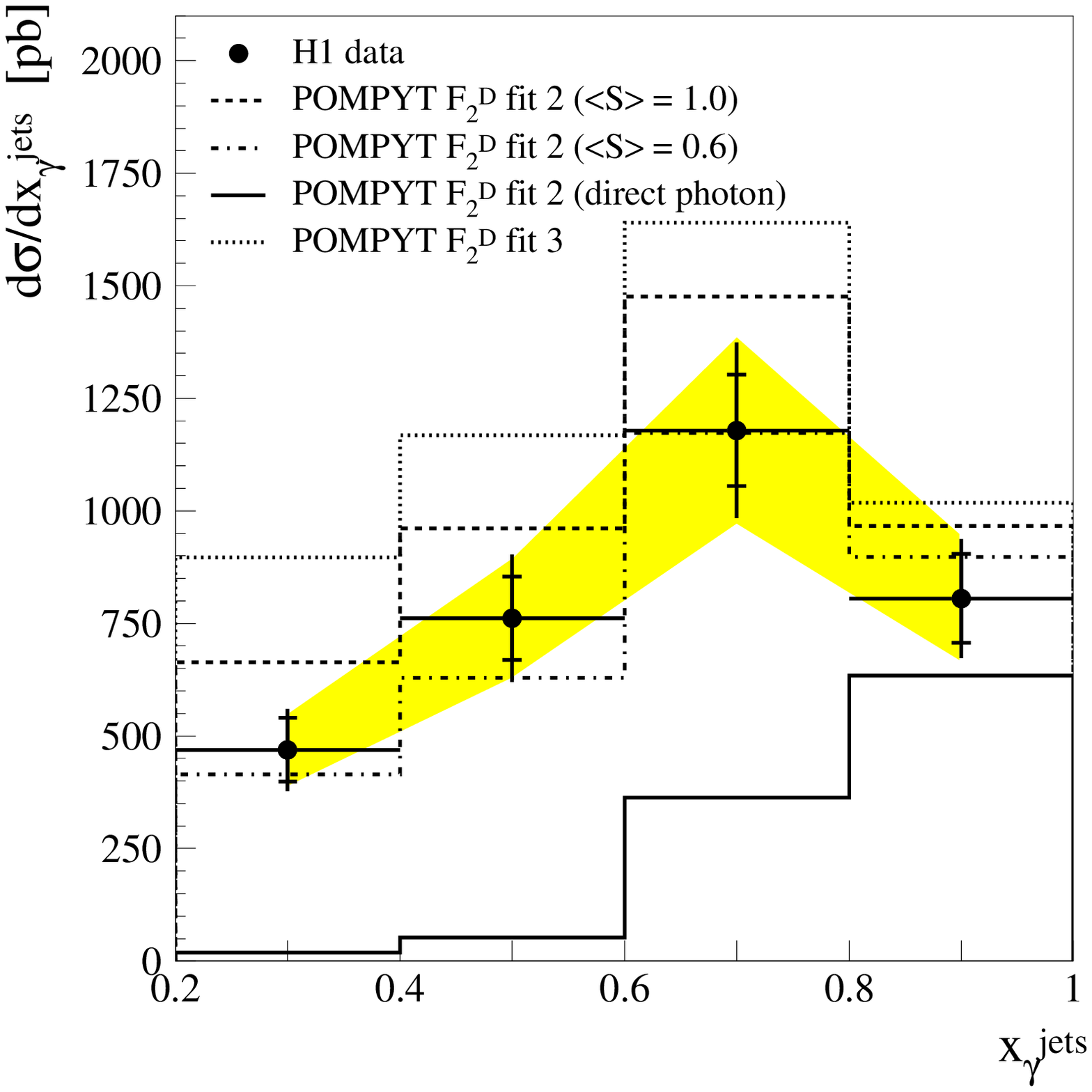,width=0.36\textwidth}}
     \put(82.5,-5){\epsfig{figure=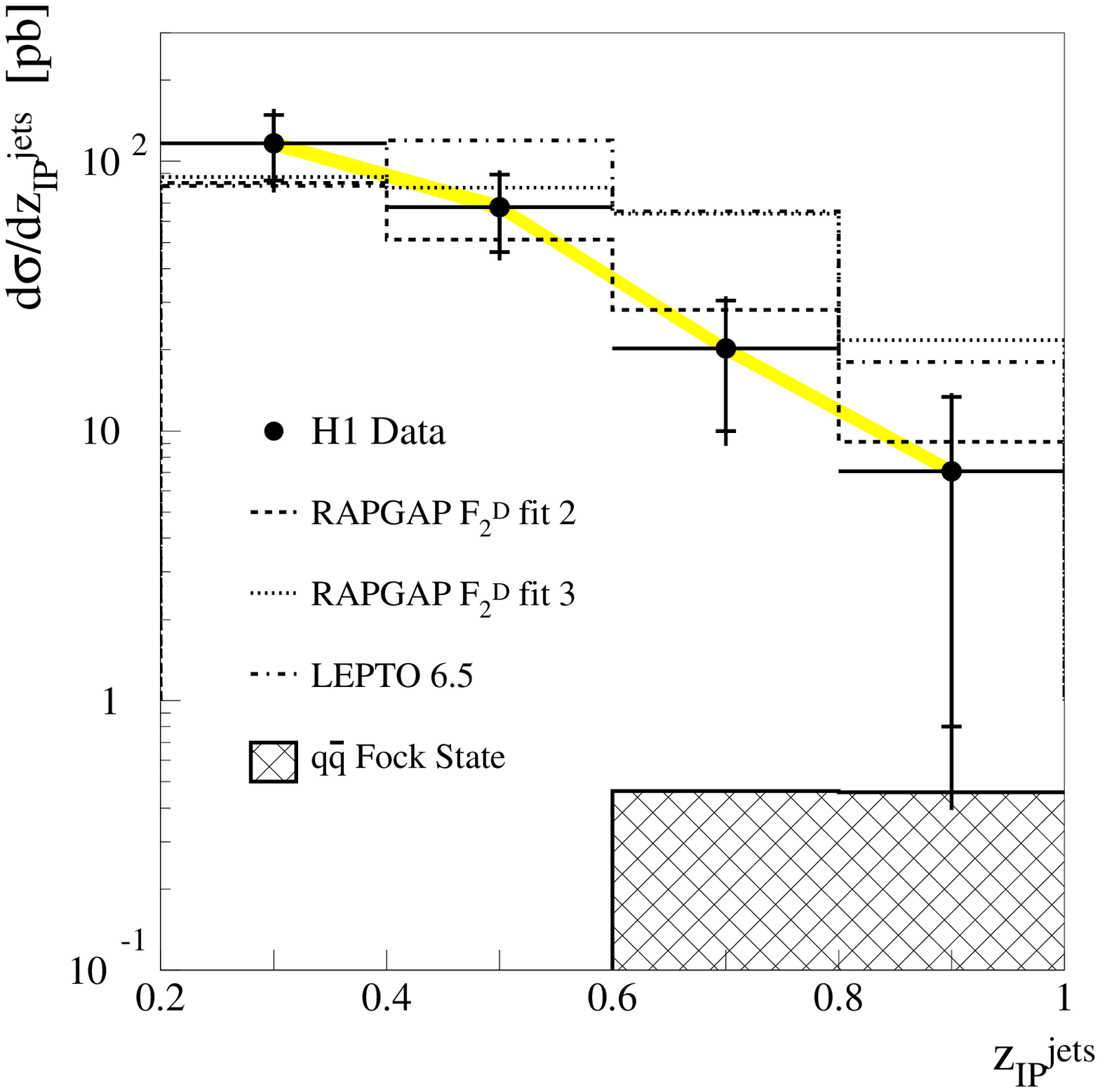,width=0.36\textwidth}}
     \put(-11,-12){\epsfig{figure=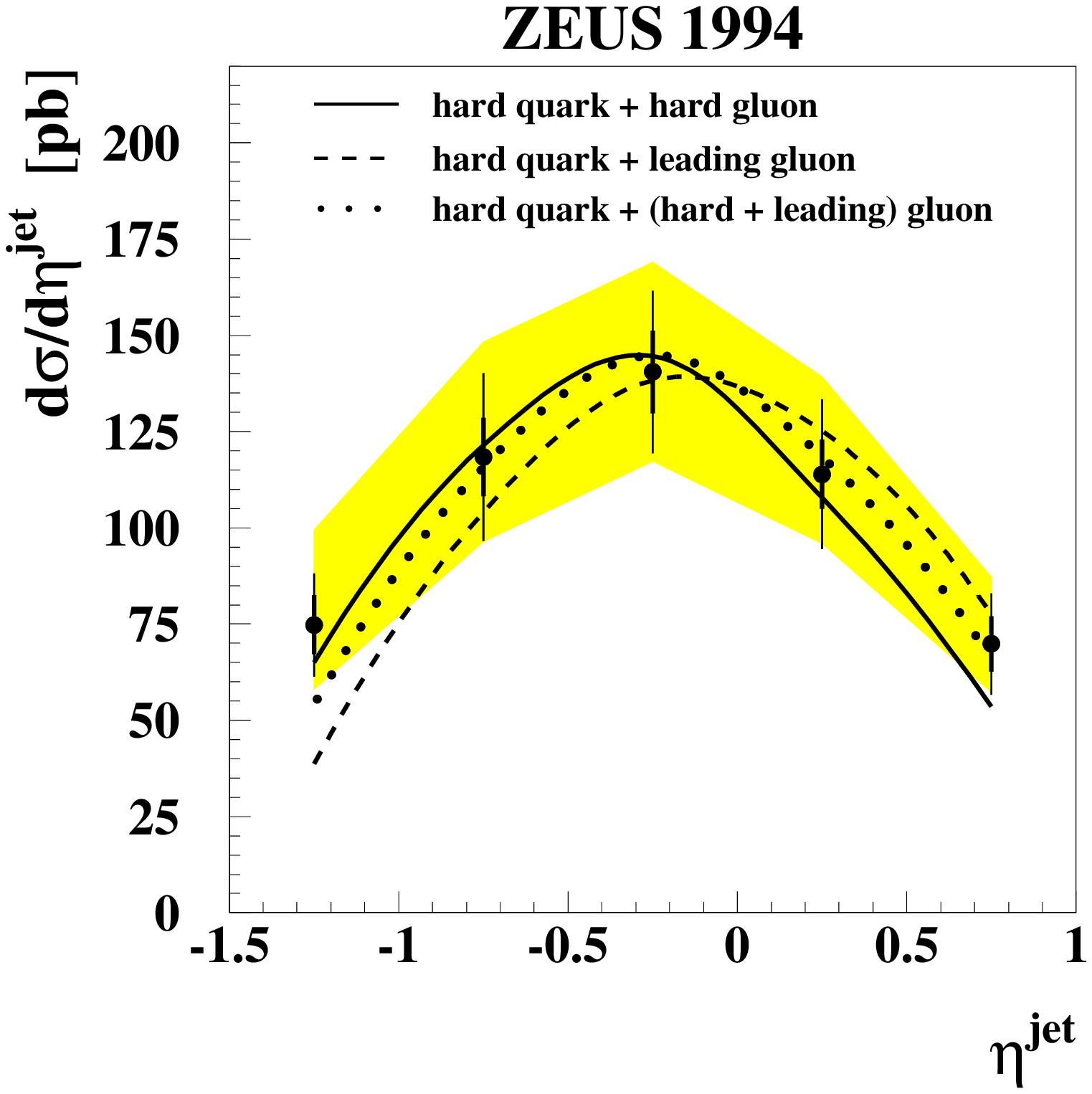,width=0.47\textwidth}}
     \put(3,2){\Large{\bf (a)}}
     \put(48,2){\Large{\bf (b)}}
     \put(90,2){\Large{\bf (c)}}
   \end{picture}
 \end{center}
 \caption {(a) The diffractive dijet photoproduction cross section differential
in pseudorapidity. The results of combined fits to the dijet data and to
a measurement of $F_2^D$, in which the pomeron has a 
gluon dominated structure, are superimposed. (b) Differential 
cross section for diffractive dijet
photoproduction as a function of $\xgamj$. The data
are compared with the predictions of the POMPYT Monte Carlo model, based on
two sets of gluon dominated pomeron parton distributions. The effect of 
possible rapidity gap destruction due to spectator interactions in resolved
photon interactions is shown by applying a constant weighting factor
$\av{S} = 0.6$ to all resolved photon interactions in the simulation. 
(c) Differential cross section for diffractive dijet
electroproduction as a function of $\zpomj$. The data
are compared to the RAPGAP model based on two gluon dominated sets of parton 
distributions, to the LEPTO model in which BGF also dominates and to a 
model \protect \cite{bartels:jets} \protect
of the exclusive production of $q \bar{q}$ final states.}
 \label{jets}
\end{figure}

Figure~\ref{jets}a shows a measurement of the pseudorapidity distribution
of diffractively produced dijets in photoproduction \cite{zeus:jets}.
A combined leading order DGLAP fit is
performed to the dijet data at a scale $E_t$
and a measurement~\cite{ZEUS:F2D93} of 
$F_2^{D(3)}$. The relatively high rate of dijet production
cannot be reproduced with a quark dominated pomeron. Good fits are obtained
with a variety of parameterisations in which a `hard' gluon distribution
dominates the pomeron structure.

Figure~\ref{jets}b shows the $\xgamj$ distribution in 
photoproduction \cite{h1:jets}. There
are clear contributions from resolved as well as direct photon interactions.
Compared through the POMPYT \cite{pompyt}
Monte Carlo model with partonic models of the
pomeron, it is clear that quark dominated 
parton distributions underestimate the dijet rates by large factors. The 
gluon dominated partons derived from the QCD fits to $F_2^D$ 
(section~\ref{pomsf})
give a fair description of the data with $\pt$
as the factorisation scale. 
In the region dominated by resolved photon
interactions ($\xgamj \lapprox 0.8$), there is some evidence for an excess
in the prediction, which may be related to rapidity
gap destruction effects \cite{survive}.

Figure~\ref{jets}c shows the $\zpomj$ distribution in 
DIS \cite{h1:jets}. There are significant contributions in the
region of large $\zpomj$, as expected for the dominance of gluons with 
large fractional momenta in the exchange. 
The data are compared through
the Monte Carlo program RAPGAP \cite{rapgap} with the 
gluon dominated pomeron parton
distributions extracted from $F_2^D$. 
Once again, 
a good description of the data is obtained
with the factorisation scale taken to
be $\pt$. A two-gluon exchange model \cite{bartels:jets}, containing
only the $q \bar{q}$ fluctuation of the photon 
fails to produce the large contributions at low $\zpomj$,
indicating that in this picture, more complex states such as $q \bar{q} g$
are required to generate the large $\pt$ and $\mx$ dijet final states.

\section{Summary}

The HERA experiments have produced a large volume of data concerned with 
diffraction and colour-singlet exchange. The kinematic range covered and
precision of data on exclusive vector meson production and virtual
photon dissociation is continually increasing.

The effective pomeron intercept describing the 
centre of mass energy dependence at fixed $\mx$ and $t = 0$ becomes larger
where hard scales such as $Q^2$ or a heavy quark mass are introduced. 
Together with numerous other observations, this motivates a perturbative
QCD based approach. For the case of exclusive vector meson
production, models based on the fluctuation of the photon into a $q \bar{q}$
pair and the subsequent exchange of a pair of gluons taken from the 
proton parton distributions successfully reproduce the enhanced energy 
dependence and other aspects of the problem. 
For the case of virtual photon dissociation,
models of this type can also describe the data, provided the fluctuation of
the photon to $q \bar{q} g$ states is also considered. 

Where large diffractive
final state masses are produced, further exchanges, in addition to the 
pomeron, are in evidence. It is possible to decompose the data
into contributions from a series of Regge
exchanges, with pomeron, $f$ and $\pi$
exchange all being significant in leading proton production. The leading 
neutron cross section
is saturated by the expectations for $\pi$ exchange. 

In models in which partonic substructure is ascribed to the exchanges, evolving
according to the DGLAP equations, the pomeron structure is found to be 
dominated by gluons carrying large fractions of the exchanged momentum. This
conclusion is confirmed through the measurement of final state observables
such as dijet rates. The leading neutron cross section has been used to extract
a structure function for the pion.

Complete analyses of the helicity structure of vector meson production 
processes
have now been made. In the high $Q^2$ regime, the cross section for $\rho$ and
$\phi$ production by longitudinal photons becomes much larger than that for
transverse photons. In this region, 
clear evidence is found for a violation of the $s$ channel helicity 
conservation hypothesis.

\section*{Acknowledgements}

Thanks to Dusan Bruncko and his team for organising such a pleasant conference,
to Richard Mara\v{c}ek for showing me the darker sides of Eastern Slovakia
and Pavel Murin for putting my life at risk in the High Tatras mountains! I
am also grateful to Martin Erdmann and Paul Thompson for proof reading this
document.




\begin{thebibliography}{99}

\bibitem{diff:review} K.\ Goulianos, {\em Phys. Rev.} {\bf 101} (1983) 169. 

\bibitem{chicago} P.\ Newman, {\em hep-ex/9707020}.

\bibitem{alexei} H1 Collab., S.\ Aid \etalk
{\em Nucl. Phys.} {\bf B472} (1996) 3. \\ 
                 H1 Collab., C.\ Adloff \etalk
{\em Zeit. Phys.} {\bf C75} (1997) 607. 

\bibitem{zeus:hight} ZEUS Collab., Conf. Paper 788, 29th Intern. Conf. on 
HEP, Vancouver, Canada (1998). 

\bibitem{zeus:gprho} ZEUS Collab., J.\ Breitweg \etalk
{\em Eur. Phys. J} {\bf C2} (1998) 247. 

\bibitem{mx:gammap} H1 Collab., C.\ Adloff \etalk
{\em Zeit. Phys.} {\bf C74} (1997) 221.

\bibitem{rho:gammap} H1 Collab., S.\ Aid \etalk
{\em Nucl. Phys.} {\bf B463} (1996) 3. 

\bibitem{zeus:mx} ZEUS Collab., J.\ Breitweg \etalk
{\em Zeit. Phys.} {\bf C75} (1997) 421.

\bibitem{low:nussinov} F.\ Low, {\em Phys. Rev.} {\bf D12} (1975) {163}. \\
                S.\ Nussinov, {\em Phys. Rev. Lett.} {\bf 34} (1975) {1286}.

\bibitem{hight} ZEUS Collab., M.\ Derrick \etalk
{\em Phys. Lett.} {\bf B369} (1996) 55. \\
                H1 Collab., paper 274 at 
Int. Europhys. Conf. on HEP, Jerusalem, August 1997. \\
                H1 Collab., paper 380 at 
Int. Europhys. Conf. on HEP, Jerusalem, August 1997. \\
                H1 Collab., Conf. Paper 570, 29th Intern. Conf. on 
HEP, Vancouver, Canada (1998). 

\bibitem{zeus:fps1} ZEUS Collab., M.\ Derrick \etalk
{\em Z. Phys} {\bf C73} (1997) 253.

\bibitem{H1:LB} H1 Collab., C.\ Adloff et al., DESY {\bf 98-169}, submitted
to {\em Eur. Phys. J. C}.

\bibitem{zeus:fnc} S.\ Bhadra \etalk {\em Nucl. Instr. and Meth} {\bf A394}
(1997) 121.

\bibitem{omega} ZEUS Collab., M.\ Derrick \etalk
{\em Zeit. Phys.} {\bf C73} (1996) 73.

\bibitem{rho:prime} H1 Collab., Conf. Paper pa01-088, 28th Intern. Conf. on
HEP, Warsaw, Poland (1996).

\bibitem{psi:prime} H1 Collab., C.\ Adloff \etalk
{\em Phys. Lett.} {\bf B421} (1998) 385. \\
                   H1 Collab., Conf. Paper 572, 29th Intern. Conf. on 
HEP, Vancouver, Canada (1998). 

\bibitem{upsilon} ZEUS Collab., J.\ Breitweg \etalk
{\em Phys. Lett.} {\bf B437} (1998) 432. \\
                  H1 Collab., Conf. Paper 574, 29th Intern. Conf. on 
HEP, Vancouver, Canada (1998). 

\bibitem{soft} A.\ Donnachie, P.\ Landshoff, 
       {\em Phys. Lett.} {\bf B296} (1992) 227. 

\bibitem{vm:gg} L.\ Frankfurt \etalk {\em Phys. Rev.} {\bf D54} (1996) 3194. \\
                A.\ Martin \etalk {\em Phys. Rev.} {\bf D55} (1997) 4329. \\
                I.\ Royen, J.\ Cudell, {\em hep-ph/9807294}. 

\bibitem{zeus:phi} ZEUS Collab., M.\ Derrick \etalk
{\em Phys. Lett.} {\bf B377} (1996) 259.

\bibitem{h1:jpsipub} H1 Collab., S.\ Aid \etalk
{\em Nucl. Phys.} {\bf B472} (1996) 3.

\bibitem{zeus:jpsi} ZEUS Collab., J.\ Breitweg \etalk
{\em Z. Phys.} {\bf C75} (1997) 215.

\bibitem{h1:jpsi} H1 Collab., Conf. Paper 572, 29th Intern. Conf. on 
HEP, Vancouver, Canada (1998).

\bibitem{stot} ZEUS Collab., M.\ Derrick \etalk
{\em Z. Phys.} {\bf C63} (1994) 391. \\
               H1 Collab., S.\ Aid \etalk
{\em Z. Phys.} {\bf C69} (1995) 27.

\bibitem{zeus:fps} ZEUS Collab., J.\ Breitweg \etalk 
        {\em Eur. Phys. J. } {C1} (1998) 81.
                   
\bibitem{H1:F2D3} H1 Collab., C.\ Adloff \etalk
          {\em Zeit. Phys.}{\bf C76} (1997) 613. 

\bibitem{H1:F2Dprelim} H1 Collab., Conf. Paper 571, 29th Intern. Conf. on 
HEP, Vancouver, Canada (1998).

\bibitem{diff:hardscat} G.\ Ingelman, P.\ E.\ Schlein, 
                    {\em Phys. Lett.} {\bf B152} (1985) 256. 

\bibitem{ariadne} L.\ L\"{o}nnblad, {\em Comput. Phys. Commun.} {\bf 71}
(1992) 15.

\bibitem{lepto} G.\ Ingelman et al., {\em Comput. Phys. Commun.} {\bf 101}
(1997) 108.

\bibitem{ZEUS:LB} ZEUS Collab., Conf. Paper 789, 29th Intern. Conf. on 
HEP, Vancouver, Canada (1998).

\bibitem{lb:regge} A.\ Szczurek et al., {\em Phys. Lett.} {\bf B428} 
(1998) 383. \\
                   B.\ Kopeliovich et al., {\em Zeit. Phys.} {\bf C73} (1996)
125. 
        
\bibitem{angular:distn} K.\ Schilling, G.\ Wolff, {\em Nucl. Phys.} 
{\bf B61} (1973) 381.

\bibitem{h1:rho} H1 Collab., Conf. Paper 564, 29th Intern. Conf. on 
HEP, Vancouver, Canada (1998).

\bibitem{SCHC} F.\ Gilman \etalk {\em Phys. Lett.} {\bf B31} (1970) 387. \\
               T.\ Bauer \etalk {\em Rev. Mod. Phys.} {\bf 50} (1978) 261.

\bibitem{zeus:angle} ZEUS Collab., addendum to Conf. Papers 792, 793, 
29th Intern. Conf. on HEP, Vancouver, Canada (1998).

\bibitem{ivanov} D. \ Ivanov, R.\ Kirschner,
       {\em Phys. Rev.} {\bf D58} (1998) 114026

\bibitem{zeus:rho} ZEUS Collab., J.Breitweg et al.,
DESY {\bf 98-107}, submitted to {Eur. Phys. J. C}

\bibitem{qqbarg} M.\ Ryskin, {\em Sov. J. Nucl. Phys.} {\bf 52} (1990) 529. \\
                 N.\ Nikolaev, B.\ Zakharov, {\em Z. Phys.} {\bf C53} (1992)
331. \\
                 M.\ W\"{u}sthoff, {\em Phys. Lett.} {\bf D56} (1997) 4311. \\
                 W.\ Buchm\"{u}ller \etalk {\em Nucl. Phys.} {\bf B487}
(1997) 283. \\
                 W.\ Buchm\"{u}ller \etalk DESY {\bf 98-113}, 
{\em hep-ph/9808454}.

\bibitem{bartels} J.\ Bartels \etalk, DESY {\bf 98-034}, {\em hep-ph/9803497}.

\bibitem{grv:pi} M.\ Gl\"{u}ck \etalk {\em Z. Phys.} {\bf C53} (1992) 651. \\
M.\ Gl\"{u}ck \etalk {\em Z. Phys.} {\bf C67} (1995) 433.

\bibitem{thrust} ZEUS Collab., J.\ Breitweg \etalk
    {\em Phys. Lett.} {\bf B421} (1998) 368. \\
                 H1 Collab., C.\ Adloff \etalk 
    {\em Eur. Phys. J.} {\bf C1} (1998) 495.

\bibitem{eflow} H1 Collab., C.\ Adloff \etalk 
    {\em Phys. Lett.} {\bf B428} (1998) 206. \\
                ZEUS Collab., Conf. Paper 787, 29th Intern. Conf. on 
HEP, Vancouver, Canada (1998).

\bibitem{multip} H1 Collab., C.\ Adloff \etalk 
    {\em Eur. Phys. J.} {\bf C5} (1998) 439.

\bibitem{zeus:jets} ZEUS Collab., J.\ Breitweg \etalk
               DESY {\bf 98-045}, submitted to {\em Eur. Phys. J.} 

\bibitem{h1:jets} H1 Collab., C.\ Adloff \etalk DESY {\bf 98-092}, 
Submitted to {\em Eur. Phys. J.}

\bibitem{charm} ZEUS Collab., Conf. Paper 785, 29th Intern. Conf. on 
HEP, Vancouver, Canada (1998). \\
                H1 Collab., Conf. Paper 558, 29th Intern. Conf. on 
HEP, Vancouver, Canada (1998).

\bibitem{ZEUS:F2D93} ZEUS Collab., M. Derrick et al., 
           {\em Zeit. Phys.} {\bf C68} (1995) 569.

\bibitem{pompyt} P.\ Bruni, G.\ Ingelman, proc. 
Int. Europhys. Conf. on HEP, Marseilles, July 1993, 595. 

\bibitem{survive} J.\ Bjorken, {\em Phys. Rev.} {\bf D47} (1993) 101. \\
                  E.\ Gotsman, E.\ Levin, U.\ Maor,
                  {\em Phys. Lett.} {\bf B309} (1993) 199. \\
                  E.\ Gotsman, E.\ Levin, U.\ Maor,
                  {\em Phys. Lett.} {\bf B438} (1998) 229.

\bibitem{rapgap} H.\ Jung, {\em Comp. Phys. Commun.} {\bf 86} (1995) 147.

\bibitem{bartels:jets} J.\ Bartels \etalk
                    {\em Phys. Lett.} {\bf B379} (1996) 239. \\
                    J.\ Bartels \etalk 
                    {\em Phys. Lett.} {\bf B386} (1996) 389.




\end{thebibliography}
\end{document}